\documentclass[pre, twocolumn,unsortedaddress, showpacs]{revtex4}
\usepackage{epsfig}
\usepackage{amssymb}

\begin{document}

\title{Multi-phase-field analysis of short-range forces between diffuse interfaces}

\author{N. Wang}
\affiliation{Physics Department and Center for Interdisciplinary Research on Complex Systems, Northeastern University, Boston, MA 02115, USA}
\author{R. Spatschek}
\affiliation{Interdisciplinary Centre for Advanced Materials Simulation (ICAMS), Ruhr-Universit{\"a}t Bochum, Germany}
\affiliation{Physics Department and Center for Interdisciplinary Research on Complex Systems, Northeastern University, Boston, MA 02115, USA}
\author{A. Karma}
\affiliation{Physics Department and Center for Interdisciplinary Research on Complex Systems, Northeastern University, Boston, MA 02115, USA}

\date{\today}

\begin{abstract}
We characterize both analytically and numerically 
short-range forces between spatially diffuse interfaces in multi-phase-field models of polycrystalline materials.
During late-stage solidification, crystal-melt interfaces may attract or repel each other depending on the degree of misorientation between impinging grains, temperature, composition, and stress. To characterize this interaction, we map the multi-phase-field equations for stationary interfaces to a multi-dimensional classical mechanical scattering problem. From the solution of this problem, we derive asymptotic forms for short-range forces between interfaces for distances larger than the interface thickness. The results show that forces are always attractive for traditional models where each phase-field represents the phase fraction of a given grain. Those predictions are validated by numerical computations of forces for all distances. Based on insights from the scattering problem, we propose a new multi-phase-field formulation that can describe both attractive and repulsive forces in real systems. This model is then used to investigate the influence of solute addition and a uniaxial stress perpendicular to the interface. Solute addition leads to bistability of different interfacial equilibrium states, with the temperature range of bistability increasing with strength of partitioning. Stress in turn, is shown to be equivalent to a temperature change through a standard Clausius-Clapeyron relation. The implications of those results for understanding grain boundary premelting are discussed.
\end{abstract}


\maketitle

\section{Introduction and summary}

Multi-phase-field models provide a powerful method to simulate complex interfacial patterns in a wide range of applications including polyphase and/or polycrystalline solidification 
\cite{Steetal96,Wheetal96,Loetal2001,Nestler,Plapp,Steinbach09}, 
grain growth \cite{FanChe97,Chen02g}, as well as domain structures and solid-state phase transformations \cite{Chen02}. The equilibrium and non-equilibrium properties of isolated interfaces in multi-phase-field models are by now well-understood. Well-developed procedures exist for selecting model parameters in order to match some experimentally specified set of interfacial free-energies and mobilities \cite{Plapp,Steinbach09}. In contrast, the interactions between interfaces has remained comparatively more poorly characterized. Interfaces in phase-field models are inherently spatially diffuse. Hence, they  interact when their distance becomes roughly comparable to the interface width $\sim \xi$. Those interactions can strongly influence  the behavior of polycrystalline materials in many processes (such as sintering and solidification) where interfaces come into close contact at various processing stages.  

There have been a few studies of 
grain coalescence using multi-phase-field models \cite{Rappaz03,NIST1} as well as frame-invariant phase-field models
with an order parameter representing the local crystal orientation \cite{LobWar02,Carter}. Those studies have yielded useful insights but have been mostly numerical due to the inherent difficulty to treat analytically the interaction between diffuse interfaces.

In this paper, we develop an analytical approach to compute short-range interactions between diffuse interfaces in multi-phase-field models. This approach is based on recasting the multi-phase-field equations for stationary interfaces in the form of a classical mechanical scattering problem. A one-dimensional mechanical analog is standard for treating the properties of isolated stationary phase-field interfaces \cite{Langer}. It has also been used to treat interactions between nonlinear fronts in the real Ginzburg-Landau equation \cite{Hakim}, which is analogous to the equation for a single phase-field.

In a multi-phase-field context, the mechanical analog becomes higher dimensional, and hence more difficult to analyze. It describes the motion of a point particle moving in $N$ dimensional space where $N$ is the number of phase fields. In standard multi-phase-field models, each phase-field $\phi_i$ describes the fraction of a given phase or grain orientation, which varies smoothly between zero and unity, with the physical constraint that $\sum_{i=1}^N \phi_i=1$. The mechanical problem is therefore subject to this constraint, but can also be formulated in $N-1$ dimensions after elimination of this constraint. This problem can be solved using conservation of total mechanical energy, which is the sum of kinetic and potential parts; the kinetic energy is related to square-gradient terms in the multi-phase-field free-energy functional and the potential energy is just the bulk free-energy density in this functional, albeit with the opposite sign. The solution yields asymptotically exact analytical expressions for the forces between interfaces for distances $W$ large compared to the interface thickness ($W\gg \xi$). 

As a concrete example of application, 
we use the mechanical analog supplemented by numerics to characterize 
the interaction of crystal-melt interfaces. This interaction is relevant for understanding grain coalescence and grain-boundary premelting phenomena. The latter has been extensively studied experimentally \cite{Glicksman72,Hsieh89,Alsayed05} and theoretically using lattice models \cite{Kikuchi80,Besold94}, atomistic molecular dynamics or Monte Carlo simulations \cite{Hoyt09,NIST2,Jinetal09} (with earlier references therein), multi-phase-field models \cite{Rappaz03,NIST1}, frame-invariant phase-field models \cite{LobWar02,Carter}, and phase-field crystal models \cite{Berry08,Plapp08}. Even though grain boundary premelting is not fully understood, there is emerging concensus  that it originates fundamentally from a repulsive interaction between crystal-melt interfaces for high-energy grain-boundaries, which is directly relevant for the present study. This repulsion gives rise to the formation of an intergranular liquid film with a width that diverges at the melting point. 
\begin{figure}
\begin{center}
\includegraphics[width=8.5cm]{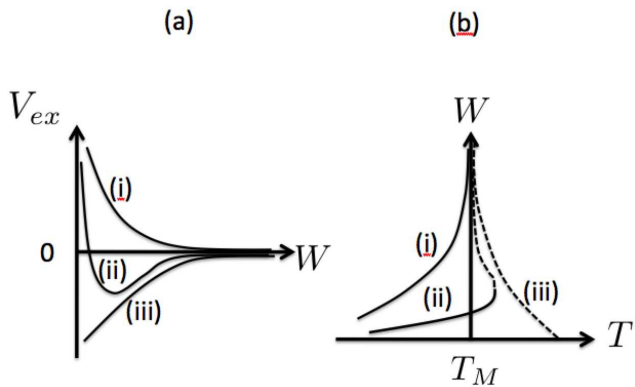}
\caption{Schematic plots of (a) disjoining potential $V_{ex}$ and (b) liquid film width $W$ versus
temperature for three qualitatively different behaviors in an elementary material (i) purely repulsive, (ii) repulsive-attractive, and (iii) purely attractive. The solid (dashed) lines in (b) denote stable (unstable) equilibrium states. The disjoining potential $V_{ex}$ represents the excess interfacial free-energy due to the interaction between interfaces (i.e., the total excess interfacial free-energy minus twice the crystal-melt free-energy) and $-dV_{ex}(W)/dW$ is the thermodynamic driving force causing interfaces to attract or repel each other. A uniaxial tensile (compressive) stress is predicted to have a similar effect as a temperature increase (decrease) in (b) through a Clausius-Clapeyron relation. The equilibrium state is monostable with a unique $W$ below the melting point in (i) but can become bistable with two different $W$ values with sufficient solute addition.}
\label{forces}
\end{center}
\end{figure}

The repulsive force responsible for the ``premelting'' of high-energy boundaries was computed in recent molecular dynamics of pure Ni \cite{Hoyt09} and a two-dimensional phase-field crystal study of hexagonal crystals \cite{Plapp08}. This force was found to decay exponentially with increasing distance between interfaces in qualitative agreement with the form traditionally assumed in sharp-interface theories \cite{Rappaz03,Widom}. In addition, for low-energy boundaries, the phase-field crystal study revealed that this force is attractive at large distance but repulsive at short distance, as also recently observed in a molecular dynamics simulation study of different grain-boundary types \cite{Jinetal09}. In this attractive-repulsive case, the force vanishes at some intermediate equilibrium liquid film thickness, which remains finite at the melting point. For two grains with the same crystal orientation, in turn, the force between crystal-melt interfaces is purely attractive, in agreement with the fact that such grains generally coalesce to form a single grain. These three qualitatively different behaviors: purely repulsive (i) for high-energy boundaries,  attractive-repulsive (ii) for low-energy boundaries, and purely attractive (iii) for grains of the same orientation are depicted schematically in Fig. \ref{forces}. 

From a practical standpoint, liquid films can lead to a significant reduction of the shear resistance of a polycrystalline mush and have been invoked recently to explain hot cracking of metallic alloys during late-stage solidification \cite{Rappaz03,Wangetal04}. Therefore the ability to reproduce the correct cross-over from attractive to repulsive behavior with increasing grain-boundary energy is essential for modeling this phenomenon, and constitutes a stringent test for continuum models of polycrytalline materials at high homologous temperature. Ideally, a multi-phase-field model should provide enough flexibility to reproduce force-distance ($-dV_{ex}(W)/dW$ versus $W$) curves with the characteristics of Fig. \ref{forces}, which can be computed from molecular dynamics simulations \cite{Hoyt09,Jinetal09}. 

A main finding of the present paper is that the standard multi-phase-field formulation \cite{Nestler,Plapp,Steinbach09} is unable to reproduce the purely repulsive behavior for high-energy boundaries, corresponding to case (i) in Fig. \ref{forces}, although it reproduces well the other behavior (ii), as well as (iii) that is a limiting case of (ii) for vanishing misorientation. In this respect, the standard multi-phase-field approach is more limited than the frame-invariant phase-field formulation that is able to reproduce all three behaviors \cite{LobWar02,Carter}. Using the mechanical analog, we show that this limitation of multi-phase-field models stems from the fact that phase fields generally represent phase (grain) fractions locally in space, and hence are constrained in the interval $0\le \phi_i\le 1$ in this interpretation.

This result appears to contradict the finding of a purely repulsive behavior for a high-energy boundary in the recent multi-phase-field study of Cu-Ag alloys by Mishin et al. \cite{NIST1}. However, these authors used a modified multi-phase-field formulation that allows some phase-fields to become negative ($\phi_i<0$) in the region where diffuse interfaces overlap. Therefore, their results do not contradict our finding that multi-phase-field models do not model pure repulsion when formulated in the traditional way where the $\phi_i$'s represent positive phase/grain fractions. Rather, when interpreted in the light of the present analysis, the study of Mishin {\it et al.} \cite{NIST1} shows that the multi-phase-field approach can be modified to reproduce all desired behaviors in Fig. \ref{forces} with a less stringent physical interpretation of the phase fields.

In the present paper, we develop a different multi-phase-field approach where the free-energy landscape is inspired from the solution of the mechanical analog problem. This approach abandons completely the interpretation of the phase-fields as phase fractions and uses a minimum number of phase-fields, as in a previous study of polyphase solidification \cite{Loetal2001}. This number is the same as the number of different grain orientations (two here for a bicrystal), which is also the number of phase fields in the standard multi-phase-field formulation after elimination of one phase-field using the constraint $\sum_{i=1}^N\phi_i=1$. The present formulation has the advantage of allowing to model the different interaction regimes in Fig. \ref{forces} by varying a parameter that controls the sign and magnitude of the interaction between interfaces
at large separation in an analytically predictable way.

This formulation is developed first for an elementary material and then extended to a dilute binary alloy to investigate analytically and numerically solute effects on interface interactions. Solute addition is found to lead to the possibility of a qualitatively different behavior. Above a threshold concentration, two equilibrium states with different widths can coexist at some temperature below the melting point, as also found in Ref. \cite{NIST1}. This temperature corresponds to a classical Maxwell point and the equilibrium state with larger (smaller) width is thermodynamically stable (metastable) above this temperature and vice versa below. We show analytically that this ``bistability'' follows from the fact that solute addition makes the long-distance interaction between interfaces more repulsive. Furthermore, we show that this effect becomes more pronounced for stronger partitioning of solute between solid and liquid. Interestingly, this type of bistability was not observed in a recent atomistic study of grain boundary premelting in Cu-Ag alloys \cite{NIST2}. In this study, the same boundary showed an attractive-repulsive behavior of type (ii) in Fig. \ref{forces} for both pure Cu and with Ag enrichment. However, this does not exclude the possibility of bistability for boundaries that already show repulsive behaviors in a pure case.

Finally, we investigate the effect of uniaxial stress perpendicular to a grain boundary on its premelting behavior. The coupling of solid-liquid phase change and stress is introduced by treating the liquid as a shear-free solid following the approach of Slutsker et al. \cite{Slutsker}. Stress is shown to be equivalent to a temperature change through a standard Clausius-Clapeyron relation for a physically plausible choice of coupling between phase field variables and elastic energy, i.e. a tensile (compressive) stress corresponds to heating (cooling). This prediction should be testable by atomistic simulations and experimentally. 

The paper is organized as follows. In the next section, we briefly review a simple sharp-interface model \cite{Rappaz03,Widom} that provides an intuitive picture of repulsive and attractive interactions. We then develop the mechanical analog in section \ref{multi}. We consider first the coalescence of two grains of the same crystal orientation that can be rigorously treated with one phase field. We then extend the approach to the more complex case of a bicrystal with three phase fields. We first discuss qualitatively why a repulsive behavior is difficult to obtain by examining the particle trajectories of the mechanical problem {\it inside} the Gibbs phase triangle and discuss how a simple triple-well one-phase-field model can produce pure repulsion. The mechanical analog is applied in section \ref{plappmod} to compute explicit asymptotic forms of interaction forces for the multi-phase-field model of Ref. \cite{Plapp}. The results confirm the qualitative analysis of particle trajectories inside the Gibbs triangle. Next, in section  \ref{attrrep}, we present our two-phase-field model of a pure bicrystal, which is a generalization of the triple-well one-phase-field model, and analyze both analytically and numerically its properties. Solute and stress effects are then treated in sections \ref{alloy}  and  \ref{elasticity}, respectively. We conclude with a few remarks in section VIII. Technical details are given in several appendices where one appendix discussed difference between double-well and double-obstacle potentials.

\section{Sharp-interface theory}
\label{sharpsec}

The simplest picture of interface interaction
is based on comparing at the melting point the excess interfacial free-energy of a dry 
grain boundary, $\gamma_{gb}$, and the excess corresponding to two well-separated solid-liquid
interfaces, $2\gamma_{sl}$. If $\gamma_{gb}>2\gamma_{sl}$, the system can in principle
lower its free-energy by forming a liquid layer, and the interfaces from two grains should repel each other.
In contrast, if $\gamma_{gb}<2\gamma_{sl}$, the interfaces should attract each other so that the
grain boundary remains dry.

This picture can be extended to predict the width $W$ of this liquid layer as a function of temperature by writing the
total excess interfacial free-energy in the form
\begin{equation}
\Delta F_{ex}=W\Delta f(T) +V_{ex}(W)+2\gamma_{sl},\label{totalexcess}
\end{equation}
where $\Delta f=f_l-f_s$ is the difference between the bulk liquid ($f_l$) and bulk solid ($f_s$) 
free-energy density and the sum of the other two terms represents the total excess interfacial free-energy.  Close to the melting temperature $T_M$,  
\begin{equation}
\Delta f(T)=L(T-T_M)/T_M, \label{deltafdef}
\end{equation}
where $L$ is the latent heat of melting per unit volume.
In addition, the quantity $V_{ex}(W)$ is the excess due to the interaction between solid-liquid interfaces, which can be assumed to have the simple form \cite{Rappaz03,Widom}
\begin{equation} \label{intro::eq1}
V_{ex}(W) = (\gamma_{gb}-2\gamma_{sl}) \exp(-W/\lambda),\label{Vexsharp}
\end{equation}
which interpolates between the limits of a dry grain boundary for $W\rightarrow 0$ and two well-separated 
solid-liquid interfaces for $W\rightarrow +\infty$. The length $\lambda$ sets the range of the exponentially decaying interaction. 
As in recent studies \cite{Plapp,Hoyt09,Jinetal09}, we refer to $V_{ex}(W)$ as the ``disjoining potential'' by analogy with the disjoining pressure of fluid physics, i.e. the derivative $-dV_{ex}/dW$ is the disjoining force that pulls interfaces a part when $\gamma_{gb}>2\gamma_{sl}$.

This form reproduces the purely repulsive and attractive cases (i) and (iii) in Fig. \ref{forces} when $\gamma_{gb}$ is larger and smaller than $2\gamma_{sl}$, respectively. 
However, it does not reproduce the intermediate behavior (ii) with short-distance repulsion and long-distance attraction predicted in recent phase-field crystal 
\cite{Plapp} and atomistic \cite{Jinetal09} modeling studies. This limitation can be attributed to the fact that Eq. (\ref{Vexsharp}) assumes sharp interfaces and does not describe the short-range repulsion associated with the formation of dislocations \cite{Plapp}, which is still present for low-energy boundaries. While both the multi-phase-field \cite{NIST1} and frame-invariant \cite{LobWar02,Carter} phase-field models also do not describe dislocations explicitly, the spatially diffuse nature of interfaces in those models suffices to produce qualitatively a short-range repulsion on a scale $\xi\sim \lambda$ and hence the intermediate behavior (ii).

The temperature dependence of the liquid layer width is obtained by minimizing the excess free-energy given by Eq. (\ref{totalexcess}) with respect to $W$, with $V_{ex}(W)$ given by Eq. (\ref{Vexsharp}). This miminization predicts a logarithmic divergence of $W$ as $T$ approaches $T_M$ from below for $\gamma_{gb}>2\gamma_{sl}$, consistent with the behavior (i) in Fig. \ref{forces}(b). For $\gamma_{gb}<2\gamma_{sl}$, it predicts that the grain boundary remains dry over a finite superheated temperature range. The dashed line (iii) in Fig. \ref{forces}(b) corresponds in this case to ``unstable'' equilibrium states. If interfaces are pulled slightly together away from their unstable equilibrium separation, they attract each other until they join in the metastable dry grain boundary state with zero width. In contrast, if they are moved slightly apart, they repel each other to form a layer width of infinite thickness.

\section{Mechanical analog}
\label{multi}
 
\subsection{Two grains of the same crystal orientation}

We consider first the coalescence of two grains with the same crystal orientation. 
A single phase-field is sufficient to distinguish between solid and liquid since both grains are equivalent. As depicted by case (iii) in Fig. \ref{forces}(b), crystal-melt interfaces are expected to attract each other for all separations $W$, since $\gamma_{gb}=0$.. As just explained at the end of the last section, this attraction implies the existence of unstable equilibrium states for $T>T_M$. The mechanical analog can be used to prove the existence of those states, and hence to conclude that the interaction is attractive.  The free-energy per unit area of interface has the form
\begin{equation}
F=\int dx\left[\frac{\sigma}{2}\left(\frac{d\phi}{dx}\right)^2+f_b(\phi,T)\right],\label{onepfF}
\end{equation}
where $f_b(\phi,T)$ is the bulk free-energy density corresponding to
a standard double-well potential with minima of equal height at $T=T_M$.  A convenient form is 
\begin{equation}
f_b(\phi,T)=f_{dw}(\phi)+g_T(\phi)L(T-T_M)/T_M\label{tilt}
\end{equation}
where $f_{dw}(\phi)=h\phi^2(1-\phi)^2$ has minima at $0$ and $1$ corresponding to liquid and solid, respectively, and $g_T(\phi)$ is a monotonously increasing function of $\phi$ with vanishing
first derivative at $0$ and $1$, and with $g_T(0)=0$ and $g_T(1)=1$.

\begin{figure}
\begin{center}
\includegraphics[width=8cm]{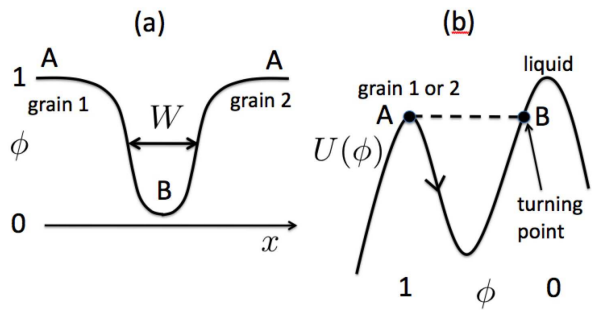}
\caption{Mechanical analog for coalescence of two grains of the same crystal orientation. 
The phase-field profile (a) correspond to the
``coordinate'' $\phi$ of a point particle moving in the ``potential'' $U(\phi)=-f_b(\phi)$ shown in (b) with $x$ (the coordinate normal to the interface) measuring ``time'' in this analogy; $f_b$ is the bulk free-energy density. The trajectory is shown for a stationary, albeit thermodynamically unstable, interface profile for $T>T_M$ where the
liquid has a lower free-energy than the solid (i.e., $f_b(0)<f_b(1)$ and hence $U(0)>U(1)$). In this case, the particle rolls down the potential energy landscape and then up to the turning point B after which it rolls back down and up to point A. This analogy also shows that a stationary interface profile cannot exist for $T<T_M$ because of the absence of turning point in this case: the particle rolls past the liquid peak and never returns. This is consistent with the fact that interfaces from two grains of the same orientation cannot repel each other.}
\label{onepf}
\end{center}
\end{figure}

The equation for planar equilibrium solutions ($\delta F/\delta \phi=0$) is 
\begin{equation}
\sigma\frac{d^2\phi}{dx^2}=-\frac{dU}{d\phi},\label{1dparticle}
\end{equation}
where we have defined $U=-f_b$. This equation has the form of Newton's law for
a one-dimensional particle of ``mass'' $\sigma$ and ``coordinate'' $\phi$ moving in a
potential $U=-f_b$, with $x$ measuring ``time''. The Hamiltonian for this dynamical system is the
total energy, which is conserved in time. It is the sum $H=K+U$ 
of the kinetic energy $K=\sigma (d\phi/dx)^2/2$ and potential energy $U$.

The proof of the existence of stationary solutions for $T>T_M$ follows immediately
form this mechanical analog. To see this, consider the phase-field profile corresponding to
an unstable equilibrium solution for $T>T_M$, which is illustrated in Fig. \ref{onepf}(a). This 
solution depicts a situation where the attractive force between the two grains due to the overlap
of the diffuse interface is balanced by the overheating that favors the liquid phase. 
In this analogy where $x$ is time, the phase-field profile corresponds to the trajectory of a particle in the
potential $U$, which has the form of a double-well potential turned up-side-down ($U=-f_b$) with the liquid
at a higher mechanical potential energy (corresponding to a lower free-energy density).
The particle leaves the equilibrium point A, corresponding to the left grain, rolls down and then up the 
potential to reach the turning point B with zero velocity, corresponding to zero slope ($d\phi/dx=0$) in the physical
phase-field profile, and then rolls back down and up to the same equilibrium point A, which
now corresponds to the right grain. It is clear that this A-B-A trajectory must exists as long as
there is a turning point, which is always true for $T>T_M$.

This mechanical analog can also be used to understand the divergence of $W$ as
the melting point is approached from above. For this, we note that the turning point
approaches the liquid-peak of the potential energy as $T$ approaches $T_M$. Therefore
the particle will spend increasingly more time close to this peak as $T$ becomes closer to
$T_M$. Therefore, this time, and hence $W$ in the analogy where time is $x$, must
diverge as $T\rightarrow T_M$.

While this picture of the divergence is only qualitative,  
a quantitative understanding   
for large $W$ is obtained by analyzing the trajectory close
to the turning point and using conservation of
mechanical energy. We sketch here the procedure and the details are elaborated 
in section \ref{plappmod}. 
Conservation of energy
implies that
\begin{equation}
H=\frac{\sigma}{2}\left(\frac{d\phi}{dx}\right)^2+U(\phi)=-L(T-T_M)/T_M,\label{cons1d}
\end{equation}
where we have used the fact that the particle has zero kinetic energy in the solid corresponding to the
stationary point A in Fig. \ref{onepf}, and thus that $H=U(1)=-f_b(1)$. Applying this conservation law at the turning
point corresponding to point B in Fig. \ref{onepf}, we obtain that
\begin{equation}
h\phi_{tp}^2\approx  L(T-T_M)/T_M,\label{turningpt}
\end{equation}
where we have used the fact that the value of $\phi$ at this point is small when
the two interfaces are well separated and that  
$g_T(\phi)$ yields a negligible contribution. This must be so because the turning point is physically located mid-way
between the two interfaces. Since the phase-field decays exponentially in space away
from the solid-liquid interfaces on both sides of this point, we would expect that 
$\phi_{tp}\approx  A\exp(-W/\lambda)$ with $\lambda\sim \xi$, where
$\xi\equiv (\sigma/h)^{1/2}$ is the interface thickness. This relation together with Eq. \ref{turningpt} 
predicts a logarithmic divergence of $W$ as $T-T_M\rightarrow 0^+$. Values for 
$A$ and $\lambda$ are easily obtained by 
matching the solutions of Eq.~(\ref{1dparticle}) in the inner region close to
the turning point and the outer regions close to the interfaces, which are both
known analytically in this simple example. 

This analysis yields an analytical expression 
for the liquid layer width as a function of temperature, 
from which one can also obtain the disjoining potential using Eq.~(\ref{totalexcess}).  
For the present example, this yields
\begin{equation}
V_{ex}(W)=-6\gamma_{sl}\exp(-W/\lambda),~(W\gg \xi),
\end{equation}
with $\lambda =\xi/\sqrt{2}$. The prefactor $6\gamma_{sl}$ is three times larger than 
predicted by the sharp-interface theory, i.e. Eq.~(\ref{Vexsharp}) with $\gamma_{gb}=0$ (but depends on the precise definition of $W$ for the diffuse interfaces). 
This is not surprising since the attractive interaction for large interface separation is
governed by properties of spatially diffuse interfaces.

\subsection{Multi-phase-field model of a bicrystal}

The standard way to describe a system consisting of a liquid and two grains of different crystal orientations with a multi-phase-field model is to use one order parameter for each grain, chosen arbitrarily here as $\phi_1$ and $\phi_2$ for grains 1 and 2, respectively, and a third ($\phi_3$) for the liquid. In addition, the constraint
\begin{equation}
\phi_1 + \phi_2 + \phi_3 = 1,  \label{multi::eq1}
\end{equation}
is imposed consistent with the interpretation that each 
$\phi_i$ represents the volume fraction of the $i^{th}$ phase. This interpretation 
also implies in principle that $0\le \phi_i\le 1$ for each phase field but those constraints are not imposed. 
The range of variation of the phase fields depends generally on the details of the free-energy functional. Standard multi-phase-field models \cite{Steinbach09} typically guarantee that $0\le \phi_i\le 1$ for all $i$. The same is true for the polyphase solidification model of Ref. \cite{Plapp}, which is adapted to a bicrystal in section \ref{plappmod}. In contrast, in the formulation of Ref. \cite{NIST1}, the $\phi_i$'s can become negative. In this subsection, we restrict our attention to using a mechanical analog to draw general qualitative conclusions about interface interactions in a broad class of models where all phase fields vary in the interval zero to unity.

The multi-phase-field free-energy functional can be written in the general form
\begin{equation} 
F = \int \, dV\, \left[f_k(\{\phi_i\}, \{\nabla \phi_i\}) +f_b(\{\phi_i\})\right],\label{multi::eq5}
\end{equation}
where $f_k$ is the ``kinetic part'' of the free-energy density that contains gradient terms and  $f_b$ is the bulk free-energy density. The former vanishes inside bulk phases while the latter remains finite.   

The stationary equations, which describe both stable and unstable equilibria, are given by
\begin{equation} \label{multi::eq3a}
\frac{\delta F}{\delta \phi_i} - \lambda_0 = 0,~{\rm for}~i=1, 2,\,{\rm and}\,3,
\end{equation}
where
\begin{equation} \label{multi::eq4}
\lambda_0 = \frac{1}{3} \sum_{i=1}^3 \frac{\delta F}{\delta \phi_i}
\end{equation}
is a Lagrange multiplier to satisfy the constraint (\ref{multi::eq1}). It is also possible to
formulate the stationary equations by using the constraint (\ref{multi::eq1}) to eliminate one
of the phase fields, chosen arbitrarily here as $\phi_3$, directly in Eq.~(\ref{multi::eq5}). The
stationary equations then have, at least formally, a simpler form without constraint
\begin{equation} \label{multi::eq3}
\frac{\delta F}{\delta \phi_i}= 0,~{\rm for}~i=1~{\rm and}~2.
\end{equation}
\begin{figure}
\begin{center}
\includegraphics[width=7cm]{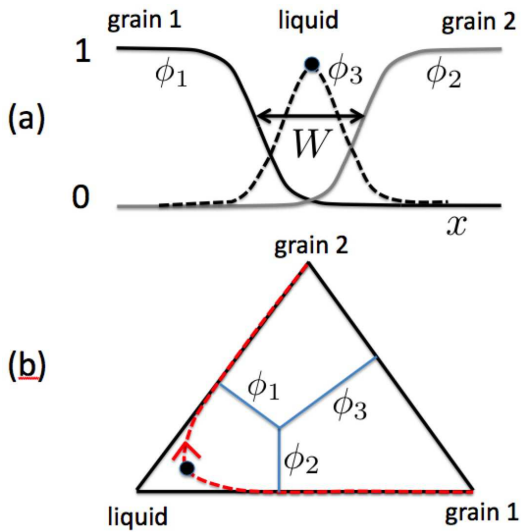}
\caption{Schematic representation of (a) the phase-field profiles for a wet bicrystal and (b) the corresponding scattering trajectory (red dashed line) inside the Gibbs phase triangle. A physically admissible phase-field profile corresponds to a scattering trajectory where a particle leaves grain 1 with zero velocity and bounces from the liquid corner to arrive at grain 2 with again zero velocity. The liquid corner is approached arbitrarily close as $T\rightarrow T_M$ and $W\rightarrow \infty$. If $T>T_M$, the liquid corner is at a higher mechanical potential energy than the corners corresponding to grains 1 or 2, and can therefore succeed to produce this large angle hard scattering. In contrast, if $T<T_M$, the liquid corner is at a lower potential energy and the particle will generally scatter at a smaller angle from the horizontal axis, thereby leaving the Gibbs triangle. The existence (absence) of trajectories for $T>T_M$ ($T<T_M$) implies that the interaction between interfaces is generally attractive at large $W$ in a multi-phase-field formulation where trajectories lie inside the Gibbs triangle.}
\label{twopf}
\end{center}
\end{figure}

For the effectively one-dimensional bicrystal geometry shown in Fig. \ref{twopf}, the stationary phase-field equations (\ref{multi::eq3}) are coupled ordinary differential equations with the independent variable $x$.
These equations are mapped to a classical mechanical problem for the motion of a particle in a conservative potential 
by introducing the generalized momenta $p_i=\partial f/\partial \dot{\phi}_i$, where we write a ``dot'' to denote $d/dx$ to emphasize the analogy to classical mechanics. From those momenta, we can construct the Hamiltonian
\begin{equation} \label{multi::eq7}
H =\sum_{i=1}^3 p_i\dot{\phi}_i-f,
\end{equation}
which is conserved in time ($\dot{H}=0$) and 
where $f=f_k+f_b$ is the total free-energy density.
Energy conservation also holds if the mechanical problem is formulated with the constraint (\ref{multi::eq1}) since the latter is holonomic, i.e. it only depends on the phase fields and not their gradients and is ``time''-independent.
Both formulations without and with constraints are shown to be completely equivalent in Appendix \ref{eliminate} and we use here the formulation without constraint as described in Table \ref{table1}.

\begin{table}
\caption{Correspondence between the multi-phase-field stationary equations describing a one-dimensional interface profile (Fig. \protect\ref{twopf}) and the classical mechanical problem of particle motion in a conservative potential where ``dot'' denotes differentiation with respect to ``time'' $x$ in this analogy.}
\begin{tabular}{c|c}
\hline\hline
Free-energy density & Lagrangian \\
$f=f_k+f_b$ & $L = T-U$ \\
\hline
Free-energy $F$ & Action $S$ \\
\hline
Position $x$ & Time $t$ \\
\hline
Phase field $\phi_i$ & Coordinate $q_i$ \\
\hline
Stationary phase field equations & Stationary action \\
$\displaystyle\frac{\delta F}{\delta \phi_i(x)}=0$ & $\displaystyle\frac{\delta S}{\delta q_i(t)}=0$ \\
\hline
Generalized momenta & Generalized momenta \\
$\displaystyle p_i = \frac{\partial f}{\partial \dot{\phi}_i}$ & $\displaystyle p_i = \frac{\partial L}{\partial \dot{q}_i}$\\[0.6ex]
\hline
Hamiltonian & Hamiltonian \\
$\displaystyle H = \sum_i p_i\dot{\phi}_i - f$ & $\displaystyle H=\sum_i p_i\dot{q}_i - L$ \\
\hline
Conservation law & Energy conservation \\
$\dot{H} = 0$ & $\dot{H} = 0$ \\
\hline\hline
\end{tabular}
\label{table1}
\end{table}

Let us now examine the particle trajectories in the bicrystal geometry of Fig. \ref{twopf}(a). As in the last subsection, the nature of the interaction for large separation ($W\gg \xi$) can be deduced from the existence of particle trajectories that correspond to physically admissible interface profiles close to melting. The interaction is attractive (repulsive) if stationary interface profiles exist for $T>T_M$ ($T<T_M$). It is useful to represent the particle trajectories in the standard Gibbs phase triangle shown in Fig. \ref{twopf}(b). The perpendicular distance of a point inside the triangle to an edge of the triangle is proportional to the volume fraction of the phase labeled at the corner opposite to this edge. Together with the constraint (\ref{multi::eq1}), this assigns a set of phase-field values $(\phi_1,\phi_2,\phi_3)$ for each point inside the triangle.  

The three corners of the Gibbs triangle correspond to minima of bulk free-energy density and hence to maxima of the conservative potential $U=-f_b$.   
Consequently, a particle trajectory that connects the two grains, shown as a dashed line in Fig. \ref{twopf}(b), leaves the grain-1 corner with zero velocity at $x=-\infty$ and ends at the grain-2 corner with zero velocity at $x=+\infty$. As $T$ approaches $T_M$, the particle must approach the liquid corner arbitrarily close and spend a long time near that corner, corresponding to a large liquid film width. 

The remaining question is whether such particle trajectories exist in a slightly undercooled and/or superheated   temperature range. For the one-dimensional mechanical analog of Fig. \ref{onepf}, the answer was clear since the point of closest approach to the liquid was a turning point. The particle only turned back if the liquid was at a higher mechanical potential energy, which required $T>T_M$ since $U=-f_b$. In the present case, the point of closest approach to the liquid (dark filled circle Fig. \ref{onepf}) is not a simple turning point since the particle has a finite velocity at this point. Instead, the liquid corner acts as a ``scattering center''. A rigorous answer to the above question therefore requires a local analysis close to the liquid corner region to solve the scattering problem that connects  
incoming and outgoing particle trajectories corresponding to diffuse solid-liquid interfaces. This analysis, described in section \ref{plappmod} for a specific choice of a multi-phase-field model, shows that scattering trajectories inside the Gibbs triangle only exist above the melting point, and hence that the interaction between interfaces is always attractive for large separation.

\begin{figure}
\begin{center}
\includegraphics[width=8cm]{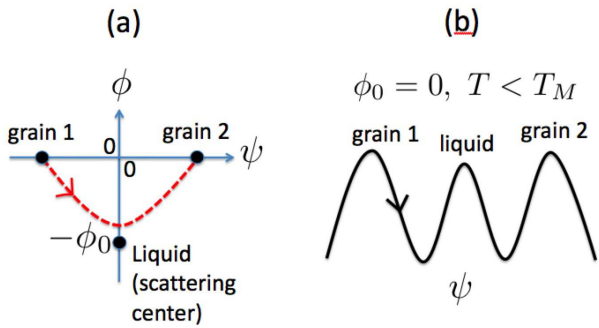}
\caption{Two-phase-field model with tunable interaction. The interaction is changed (a) by varying the distance $\phi_0$ of the liquid free-energy minimum from the axis passing through the two solid minima. The scattering becomes ``softer'', with the interaction switching from attractive to repulsive, as $\phi_0$ is decreased. In the case $\phi_0=0$ (b), the particle trajectory becomes simply one-dimensional and hops over the liquid minimum without being scattered. This trajectory can  clearly only exist for $T<T_M$ since $U=-f_b$, showing that the interaction between interfaces is repulsive in this case. }
\label{scattering}
\end{center}
\end{figure}

This answer can be qualitatively understood from the structure of the scattering problem with the help of Fig. \ref{twopf}. If $T>T_M$, the liquid corner is at a higher mechanical potential energy than the corners corresponding to grains 1 or 2, and can therefore succeed to scatter the particle at a large angle back towards the grain-2 corner. In contrast, if $T<T_M$, the liquid corner is at a lower potential energy and the particle will generally scatter with a smaller angle, thereby leaving the Gibbs triangle. The existence (absence) of trajectories for $T>T_M$ ($T<T_M$) implies that the interaction between interfaces is generally attractive at large $W$ in a multi-phase-field formulation where trajectories lie inside the Gibbs triangle.

This qualitative picture suggests how to construct a multi-phase-field approach to reproduce both attractive and repulsive interactions by relaxing the constraint that the trajectories lie inside the Gibbs triangle. The idea, which abandons the interpretation of the phase fields as volume fractions, is to construct a free-energy landscape where the free-energy density minima corresponding to the grains and the liquid are arranged in such a way that the scattering angle of the particle from the liquid corner can be tuned to change the sign of the interaction, which is attractive for ``hard'' back scattering but repulsive for ``soft'' forward scattering from grain 1 to grain 2. A model with two phase fields $\phi$ and $\psi$ built on this idea is shown schematically in Fig. \ref{scattering}, and presented in more detail in section \ref{attrrep}. This model makes it possible to continuously change the interaction from attractive to repulsive by reducing the distance $\phi_0$ of the liquid free-energy minima from the axis passing through the other two solid minima. Reducing this distance reduces the scattering angle that vanishes for $\phi_0=0$. In this extreme case, the liquid minima lies along the same  axis as the two solid minima. Therefore the particle trajectory becomes simply one-dimensional and hops over the liquid minimum without being scattered. Such a trajectory can clearly only exist for $T<T_M$ since $U=-f_b$. This rigorously proves that the interaction between interfaces is repulsive in this limit of the model.

\section{Analysis of multi-phase-field models}
\label{plappmod}

In this section, we use the mechanical analog to compute analytically the large-distance interaction between interfaces in standard multi-phase-field formulations where the particle trajectories lie inside the Gibbs phase triangle, as shown in Fig. \ref{twopf}. As discussed in the last section,
this requires an analysis of the trajectory near the liquid scattering center, which corresponds to the liquid region between the two grains. We illustrate here this computation for the specific choice of the model of Ref. \cite{Plapp}, but also consider other multi-phase-field formulations at the end of this section. The model of Ref. \cite{Plapp} has two advantages for the present analysis. First, in the simplest case of equal interfacial energies, the phase field profiles are known analytically. Second, isolated interfaces between two phases, referred hereafter as binary interfaces, run exactly along the edges of the Gibbs triangle. Therefore the interface along a given edge does not contain a spurious admixture of the phase labeled at the corner opposite to this edge ($\phi_3=0$ everywhere along the interface between grain 1 and grain 2, etc). We supplement our analysis by exact numerical computations of the forces for arbitrary distances between interfaces.


\subsection{Free-energy functional}

The individual contributions to the free-energy are
\begin{equation} \label{multi::eq8}
f_{dw} = h\sum_{i=1}^3 \phi_i^2(1-\phi_i)^2,
\end{equation}
where $h$ is a measure for the barrier height between the bulk states with the dimension of an energy density.
The gradient energy is
\begin{equation} \label{multi::eq9}
f_{k} = \frac{\sigma}{2} \sum_{i=1}^3 (\nabla\phi_i)^2,
\end{equation}
where the parameter $\sigma$ plays the role of the mass in the mechanical picture.
For the phase field model, $h$ and $\sigma$ specify the solid-liquid free-energy and the interface thickness, see below.
To allow for unequal solid-liquid and grain boundary interfacial energies, we add a grain boundary energy term which raises the free-energy well between the solid phases,
\begin{equation} \label{multi::eq10}
f_{gb} = ha\phi_1^2\phi_2^2 (2\phi_1\phi_2 + 3\phi_3 + b\phi_3^2)
\end{equation}
where only the dimensionless number $a$ influences the ratio $\gamma_{gb}/\gamma_{sl}$, and $b$ raises the free-energy bump only in the center of the Gibbs triangle but not along its boundary.
We also introduce a coupling term
\begin{equation} \label{multi::eq11}
f_c = L\frac{T-T_M}{T_M} g_T(\{\phi_i\})
\end{equation}
with the melting temperature $T_M$, the latent heat $L$ and a thermal coupling function
\begin{eqnarray}
g_T &=& - \frac{\phi_3^2}{4} \Big[ 15(1-\phi_3)(1+\phi_3-(\phi_2-\phi_1)^2) \nonumber \\
&& + \phi_3(9\phi_3^2-5) \Big]. \label{multi::eq12}
\end{eqnarray}
Then the total free-energy density is $f=f_{dw}+f_{gb} + f_{k}+f_c$, which is symmetric under exchange of $\phi_1$ and $\phi_2$ as they represent the same solid phase only in different orientations.
Hence $f_b=f_{dw}+f_{gb}+f_c$ in the above notation.
The stationary equations are given by Eqs.~(\ref{multi::eq3}) after elimination of $\phi_3$ using the constraint that the sum of the phase fields equals unity.

\subsection{Liquid film width}

We now present a method to analyze the interaction between diffuse interfaces analytically and compare the findings to the numerical results.

\begin{figure}
\begin{center}
\includegraphics[width=8.5cm]{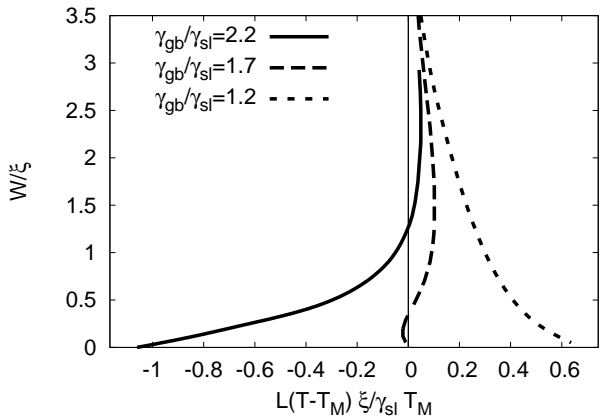}
\caption{Numerically computed liquid film width versus dimensionless overheating for $b=2$. The parts of the curves with negative slope are unstable for fixed temperature. 
The liquid layer thickness is defined as the distance between the point where the ``solid'' phase-fields cross the value $1/2$.}
\label{multi::fig2}
\end{center}
\end{figure}

In the vicinity of the liquid corner, we can linearize the phase field equations and obtain
\begin{equation} \label{multi::eq13}
\sigma\ddot{\phi}_i=2h\phi_i, \quad i=1,2,
\end{equation}
where we have eliminated the third field.
We note that the equations for both fields naturally decouple and have the same coefficients.
This property will be discussed below in a more general context and become more transparent there.
The total energy becomes in quadratic approximation
\begin{equation} \label{multi::eq14}
H = -2h (\phi_1^2 + \phi_2^2 + \phi_1\phi_2) + \sigma ( \dot{\phi}_1^2 + \dot{\phi}_2^2 + \dot{\phi}_1\dot{\phi}_2) + L\frac{T-T_M}{T_M}.
\end{equation}
On the other hand, the energy can be obtained from the limit $x\to\pm\infty$, where it is $H=0$.
The general solution of the linearized equations of motion is
\begin{equation} \label{multi::eq15}
\phi_i=C_{i1}\exp(\lambda x) + C_{i2}\exp(-\lambda x)
\end{equation}
with $\lambda = (2h/\sigma)^{1/2}$.
Symmetry with respect to exchange of the solid fields demands $C_{11}=C_{22}$ and $C_{12}=C_{21}$, and then we obtain
\begin{equation} \label{multi::eq16}
H = -4 h (4C_{11}C_{12} + C_{11}^2 + C_{12}^2) + L\frac{T-T_M}{T_M} = 0.
\end{equation}
For $T<T_M$ one of the coefficients has to become negative, but this immediately implies that the phase field coordinates will become negative at some moment.
Since this contradicts the fundamental assumption that all phase fields have to stay in the range $0$ to $1$, this shows that a solution with very wide liquid layer cannot exist below the melting temperature.
Since we know that a solution must exist which connects to the macroscopic equilibrium solution $W=\infty$ for $T=T_M$, this (unstable) branch of solution must be located above the melting temperature.
Notice that there all coefficients can be positive, and we therefore cannot exclude the existence of solutions.
This is also illustrated in Fig. \ref{multi::fig2} for $b=2$, which we obtained from the numerical solution of the stationary phase field equations that were solved by a shooting method. 
Details of the solution procedure are described in Appendix \ref{shooting}.
Here we use the expression for the solid-liquid free-energy density
\begin{equation} \label{multi::eq17}
\gamma_{sl} = \frac{\sqrt{2\sigma h}}{3},
\end{equation}
and the grain boundary energy \cite{Plapp}
\begin{equation} \label{multi::eq18}
\gamma_{gb} = 2\sqrt{2\sigma h} \int_0^1 p(1-p)\sqrt{1+ap(1-p)}\, dp.
\end{equation}
In particular we see that the interfaces asymptotically attract each other, irrespective of the value of the grain boundary energy.

\begin{figure}
\begin{center}
\includegraphics[width=6cm]{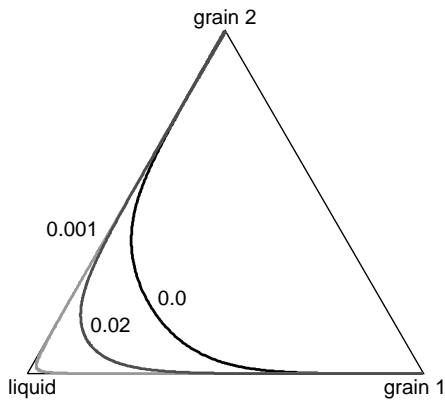}
\caption{Numerically computed trajectories in the Gibbs phase triangle for different values of the overheating $L(T-T_M)/h T_M$, which are labeled for each trajectory. The parameters
$\gamma_{gb}/\gamma_{sl}=2.2$ and $b=2$ are the same as in Fig. \protect\ref{multi::fig2}.}
\label{multi::fig1}
\end{center}
\end{figure}
For short distances $W$, however, the grains can also have a repulsive interaction, which of course does not contradict  the asymptotic prediction.
At sufficiently low temperatures, the melt layer disappears, and the solution continues as a dry branch $W=0$ towards stronger undercoolings.
The existence of these additional dry branches, which are not shown in Fig. \ref{multi::fig2} is a specific property of the model of Ref. \cite{Plapp}, and related to the absence of third phase contributions in a binary interface.

For this model, the solution first runs nearby the edge of the Gibbs triangle which connects one solid phase with the liquid phase (see Fig.~\ref{multi::fig1}).
This is not mandatory for a general model, as many phase field models have third phase contributions in a binary interface, which implies that the trajectory deviates from the edge of the Gibbs triangle. We briefly discuss this case below.
However, in the case of an interface between one grain and the liquid phase it is of course desirable not to have a contribution of the other grain in the transition region, and much care was spent on fulfilling this requirement in the above model \cite{Plapp}.

\begin{figure}
\begin{center}
\includegraphics[width=8.5cm]{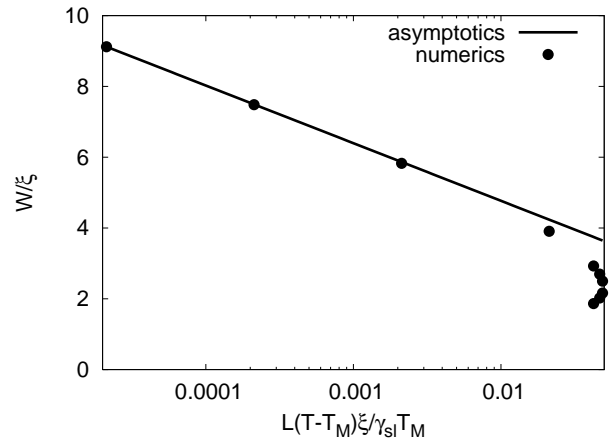}
\caption{Semi-logarithmic plot of the liquid film width as function of dimensionless overheating. The numerically computed values (solid circles) are compared to the analytical prediction Eq.~(\ref{multi::eq20}). We use $b=2$ and $\gamma_{gb}/\gamma_{sl}=2.2$ here.}
\label{multi::fig3}
\end{center}
\end{figure}

Since the binary solid-liquid interface profile is known analytically here, we can construct the solution for the asymptotic behavior for $T\to T_M^+$:
Approaching from $x=-\infty$, let the solid-liquid interfaces be located at $x_0=\pm W/2$ with a separation $W/\xi\gg 1$, where 
\begin{equation}
\xi=\left[\frac{\sigma}{h}\right]^{1/2},
\end{equation}
is the thickness of an isolated interface with profile
\begin{equation} \label{multi::eq19}
\phi_{1/2} = \frac{1}{2} \left( 1 \pm \tanh\frac{x\pm x_0}{\sqrt{2}\xi} \right).
\end{equation}
For $0\gg x\gg -x_0$ we can match its asymptotic behavior, $\phi_1(x)\simeq\exp[-\sqrt{2}(x-x_0)/\xi]$ to the result from the linearization and obtain $C_{12}=C_{21}=\exp[-W/\sqrt{2}\xi]$.
On the other hand, the inner solution has to match asymptotically a trajectory that passes along the edges of the Gibbs triangle for $T\to T_M^+$.
This implies that $\phi_1/\phi_2\to 0$ for $x\to\infty$ and therefore $C_{11}=0$.
Using the energy conservation (\ref{multi::eq16}) gives then $4hC_{12}^2= L(T-T_M)/T_M$, which together with the above finding leads to the asymptotic behavior of the unstable branch
\begin{equation} \label{multi::eq20}
W = -\left( \frac{\sigma}{2h} \right)^{1/2} \ln \left[ \frac{L}{4h} \frac{T-T_M}{T_M} \right]\quad\textrm{for}\; T\to T_M^+.
\end{equation}
The comparison to the analytical prediction is shown in Fig.~\ref{multi::fig3}.
Here we see explicitly that the liquid layer thickness diverges logarithmically when the melting point is approached, in agreement with the predictions of lattice models \cite{Kikuchi80} and molecular dynamics
simulations \cite{Hoyt09}.

\subsection{Disjoining potential}

\begin{figure}
\begin{center}
\includegraphics[width=8.5cm]{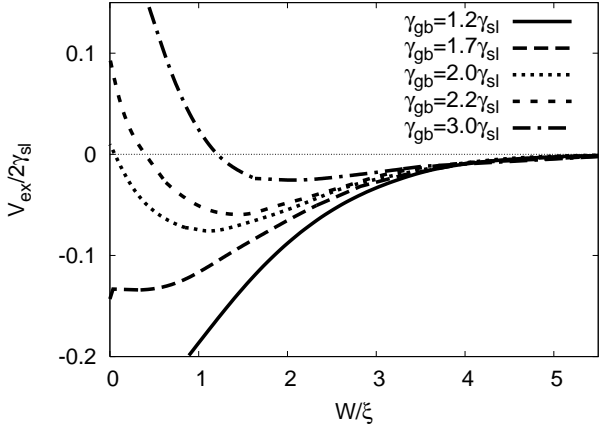}
\caption{Numerically computed plots of disjoining potential versus liquid film width for $b=2$ showing that the interaction is always attractive at large distance for the model of Ref. \protect\cite{Plapp}.}
\label{multi::fig4}
\end{center}
\end{figure}

\begin{figure}
\begin{center}
\includegraphics[width=8.5cm]{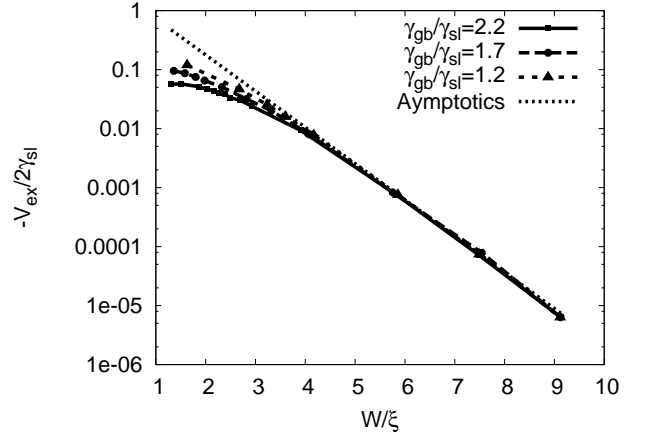}
\caption{Semi-logarithmic plot of the excess potential as function of the liquid layer thickness $W$. The data is compared to the analytical prediction Eq.~(\ref{multi::eq24}). We use $b=2$ here.}
\label{multi::fig5}
\end{center}
\end{figure}

We can obtain the disjoining potential, 
using Eq.~(\ref{totalexcess}), which yields
\begin{equation} \label{multi::eq21}
V_{ex}(W) = \Delta F_{ex}(W) + \frac{L(T-T_M)}{T_M} W - 2\gamma_{sl},
\end{equation}
where $\Delta F_{ex}(W)$ is the total excess free-energy, i.e. the total free-energy of the system minus the free-energy of a bulk solid phase occupying the same volume. The latter quantity is easily obtained by substituting the numerically computed phase-field profiles into the free-energy functional. The results plotted in Fig \ref{multi::fig4}, confirm the analytical prediction that the interaction is always attractive for large $W$. The disjoining potential can also be predicted analytically by using the fact that $d\Delta F_{ex}(W)/dW=0$ for a stationary interface, which yields the relation
\begin{equation} \label{multi::eq22}
V_{ex}'(W) = \frac{L(T-T_M)}{T_M}.
\end{equation}
This relation reflects the fact that the grain attraction is compensated 
by the overheating.
We therefore obtain by comparison with Eq.~(\ref{multi::eq20})
\begin{equation} \label{multi::eq23}
W \simeq -\left( \frac{\sigma}{2h} \right)^{1/2} \ln \left[ \frac{V_{ex}'(W)}{4h}\right],
\end{equation}
which can be solved for $V_{ex}'$ and integrated to yield
\begin{equation} \label{multi::eq24}
V_{ex}(W) \simeq -6\gamma_{sl} \exp\left( -\frac{\sqrt{2}W}{\xi} \right)\quad\textrm{for }W\to\infty.
\end{equation}
Here it becomes apparent that the grain boundary energy is not relevant for the long-range attraction.
Instead, the strength of the interaction is solely set by $6\gamma_{sl}$, in contrast to the simple model (\ref{intro::eq1}).
As anticipated, the exponential decay takes place on the scale of the interface thickness $\xi$.
This expression is compared to the numerical results in Fig. \ref{multi::fig5}, showing an excellent agreement.
\subsection{Other multi-phase-field formulations}

Let us now briefly examine other multi-phase-field models than the one of Ref. \cite{Plapp} with a more 
general expression for the free-energy functional.
We assume that the liquid phase-field is directly eliminated, so we do not need the Lagrange multiplier.
Then the potential part of the free-energy is in quadratic approximation around the liquid point $\phi_1,\phi_2=0$:
\begin{equation}  \label{multi::eq25}
f_b(\phi_1, \phi_2) = \frac{a}{2} ( \phi_1^2 + \phi_2^2 ) + b\phi_1\phi_2 + \Delta f,
\end{equation}
and the kinetic energy is
\begin{equation}  \label{multi::eq26}
f_k = \frac{c}{2} [ (\dot{\phi}_1)^2 + (\dot{\phi}_2)^2 ] + d \dot{\phi_1}\dot{\phi}_2.
\end{equation}
Terms like $\phi_i\dot{\phi}_j$ cannot appear because they violate inversion symmetry.
Positive definiteness requires $a>|b|$, and $c>|d|$, since the bulk liquid should be a stable solution;
the exchange symmetry $\phi_1 \leftrightarrow\phi_2$ is reflected by the above expressions.
Notice that the above form of the kinetic energy contains also a case that is widely used in the literature (see e.g. \cite{Nestler} and references therein)
\begin{equation} \label{multi::eq27}
\tilde{f}_{grad} = \sum_{i,j=1}^3 \gamma_{ij} (\phi_i\nabla \phi_j - \phi_j\nabla\phi_i)^2,
\end{equation}
where the coefficients $\gamma_{ij}=\gamma_{ji}$ are related to the interfacial free-energies.
In the vicinity of the liquid fixpoint they again reduce to terms of the above type.

The equations of motion are therefore
\begin{eqnarray}
c\ddot{\phi}_1 + d\ddot{\phi}_2 &=& a\phi_1 + b\phi_2, \label{multi::eq28} \\
d\ddot{\phi}_1 + c\ddot{\phi}_2 &=& b\phi_1 + a\phi_2, \label{multi::eq29}
\end{eqnarray}
We can define $\lambda_1=\sqrt{(a+b)/(c+d)}$ and $\lambda_2=\sqrt{(a-b)/(c-d)}$.
The general solution is then
\begin{eqnarray}
\phi_1 &=& c_1\exp(\lambda_1x) + c_2\exp(-\lambda_1x) \nonumber \\
&+& c_3\exp(\lambda_2x) + c_4\exp(-\lambda_2x) \label{multi::eq30} \\
\phi_2 &=& c_1\exp(\lambda_1x) + c_2\exp(-\lambda_1x) \nonumber \\
&-& c_3\exp(\lambda_2x) - c_4\exp(-\lambda_2x) \label{multi::eq31}
\end{eqnarray}
Symmetry requires $c_2=c_1$ and $c_4=-c_3$, thus we have
\begin{eqnarray}
\phi_1 &=& 2c_1\cosh(\lambda_1 x) + 2c_3\sinh(\lambda_2 x), \label{multi::eq32}\\
\phi_2 &=& 2c_1\cosh(\lambda_1 x) - 2c_3\sinh(\lambda_2 x). \label{multi::eq33}
\end{eqnarray}
Then the Hamiltonian becomes
\begin{equation} \label{multi::eq34}
H = -\Delta f -4(a+b) c_1^2 + 4(a-b)c_3^2 = 0, 
\end{equation}
where the tilt term $\Delta f$ corresponds as before to a deviation from the melting temperature ($\Delta f<0$ for $T>T_M$).

We can now distinguish three cases: $\lambda_1>\lambda_2$, $\lambda_1<\lambda_2$ and $\lambda_1=\lambda_2$.

First, if $\lambda_1<\lambda_2$, the ``even'' mode associated with $\cosh$ has the slowest decaying exponential.
In the mechanical analog picture, it corresponds to a reflection of the particle at the ``liquid'' potential hill.
Notice that the matching constants $c_i$ behave as $c_i=\tilde{c}_i\exp(-\lambda_i W/2)$, with $\tilde{c}_i$ being a number of order unity, which is determined from the matching of a single solid-liquid interface.
This shows that in the first case $c_2$ is exponentially small in comparison to $c_1$ in the limit $W\to\infty$ ($T\to T_M$).
Then this term does not appear in the energy conservation in this limit.
Since $a+b>0$ the condition $H=0$ (which follows from the fact that the energy in the pure solid is zero) can only be fulfilled for negative $\Delta f$, i.e. $T>T_M$, so the model is attractive at long distances.

Second, for $\lambda_2<\lambda_1$, only the ``odd'' (repulsive) mode survives, so now $c_1$ is exponentially small compared to $c_2$, and we can ignore the $\cosh$ part in the general solution.
Then, however, the phase fields must become negative, which is forbidden.
This mode corresponds to a particle that traverses the liquid bump, and therefore leaves the Gibbs phase triangle.

Notice that in the case of unequal decay rates a pure binary interface cannot be free of third phase contributions.
If we assume that the solid with $\phi_1=1$, $\phi_2=0$ for $x\to-\infty$ is in equilibrium with the melt $\phi_3=1$ for $x\to\infty$, all growing exponentials must be suppressed, $c_1=c_3=0$, in the above general solution (\ref{multi::eq30}, \ref{multi::eq31}).
Then, however, $\phi_2=\pm \phi_1$ in the vicinity of the liquid fixpoint, which implies that a contribution of the other solid field $\phi_2$ is always present.
The equality of the exponentials is therefore a necessary condition for the absence of third phase contributions, as it is the case for the model above \cite{Plapp};
however, it is not a sufficient condition, and a counterexample is the model \cite{Nestler}, which is based on the kinetic energy expression (\ref{multi::eq27}).

Finally, the model of Plapp and Folch \cite{Plapp} is a prototype of the last case of equal exponentials, $\lambda_1=\lambda_2$. We refrain here from performing a detailed general analysis of this case. Nonetheless we conclude that the mechanical analog, together with the restriction that the phase fields remain inside Gibbs triangle, poses a severe constraint. Therefore it is generally difficult to construct models that exhibit a long-range repulsion
when the phase fields are interpreted as phase fractions. 

\section{Tunable interaction model}
\label{attrrep}

In this section, we present a simple two-phase-field model constructed around the idea that the large-distance interaction can be made repulsive by making the scattering trajectory of the particle softer in the mechanical analog, as discussed at the end of section \ref{multi} and illustrated in Fig. \ref{scattering}.

\subsection{Model formulation}

\begin{figure}
\begin{center}
\includegraphics[angle=-90, width=9.5cm, trim=40mm -1.5cm 20mm 0cm]{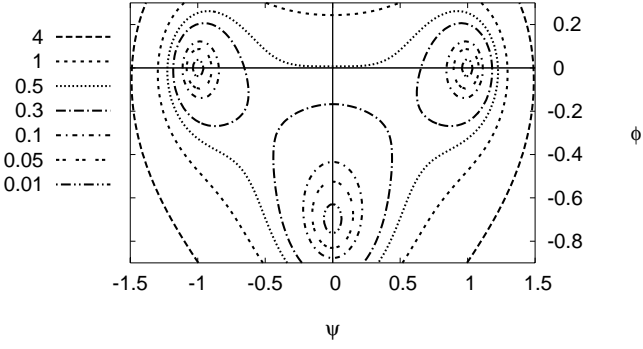}
\caption{Contour plot of the free-energy landscape for $\phi_0=0.7$, $h=1$. The  ($\psi=\pm1$,$\phi=0$) minima correspond to the two different crystal orientations and the ($\psi=0$,$\phi=-\phi_0$)  minimum to the liquid. For $\alpha=1$ the model is repulsive for $\phi_0<0.69$, in which case the scattering angle in the mechanical analog is sufficiently reduced for a particle trajectory to connect the two grains below the melting temperature.}
\label{attrrep::fig1}
\end{center}
\end{figure}

\begin{figure}
\begin{center}
\includegraphics[width=8.5cm]{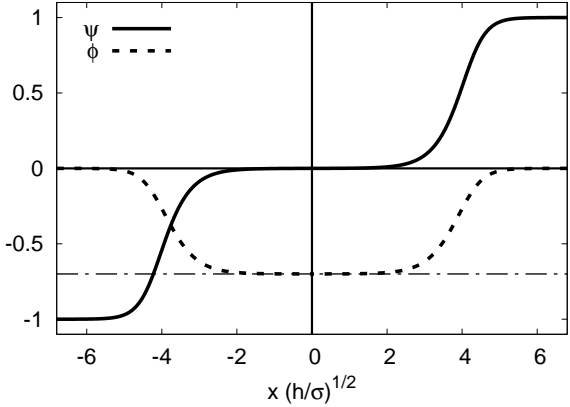}
\end{center}
\caption{Phase field profiles for a one-dimensional wetted bicrystal geometry with a liquid layer sandwiched between two grains. Here $\alpha=1$ and $\phi_0=0.7$.}
\label{attrrep::fig2}
\end{figure}

A simple polynomial form of free-energy density with the structure of Fig. \ref{scattering} is given by
\begin{equation} \label{attrrep::eq1}
f_b = h[(\psi+1)^2+\phi^2]\cdot[(\psi-1)^2+\phi^2]\cdot[\psi^2+(\phi+\phi_0)^2],
\end{equation}
where $\phi_0>0$ measures the distance of the liquid minimum at $(-\phi_0, 0)$ from the axis passing through the two solid minima at $(\psi,\phi)=(-1, 0)$, $(1,0)$. A numerical example of the free-energy landscape is given in Fig.~\ref{attrrep::fig1} %
and typical one-dimensional phase field profiles for a wet bicrystal are shown in Fig.~\ref{attrrep::fig2}. In this model, the order parameter $\psi$ has different values in the two grains of different crystal orientations ($\psi=\pm 1$) and the liquid ($\psi=0$), while $\phi$ only has different values in solid ($\phi=0$) and liquid ($\phi=-\phi_0$). Therefore, it would be tempting to loosely  interpret $\psi$ as a local measure of average crystal orientation, which vanishes in the liquid, and $-\phi/\phi_0$ as a liquid fraction that varies from zero in the solid to unity in the liquid. However, such an interpretation has to be taken with caution for several reasons. Firstly, the model is not frame-invariant, hence the interpretation of $\psi$ as a measure of local crystal orientation is not well-defined. Secondly, for $\phi_0=0$, all the free-energy minima lie on the $\psi$ axis and $|\psi|$ can equally well represent a liquid fraction in this case. Thirdly, changing $\phi_0$ has the same effect as changing the grain boundary energy and hence the misorientation. For these reasons, it is better to think as $\phi$ and $\psi$ as the minimum set of two phenomenological order parameters necessary to construct a free-energy landscape with the desired properties.

The gradient term (kinetic energy) is given by
\begin{equation} \label{attrrep::eq2}
f_k = \sigma \left[ \frac{1}{2} (\nabla\phi)^2 + \frac{\alpha}{2} (\nabla\psi)^2 \right],
\end{equation}
where we introduce $\alpha>0$ as additional parameter.
In the mechanical analog, this corresponds to a tuning of the masses.
Notice that we use the same parameters $h$ and $\sigma$ as before;
however, neither the interface width nor the interfacial free-energy can here be calculated explicitly, since the phase field profiles are not known analytically.

Finally, to favor the liquid or solid states, we introduce a thermal tilt,
\begin{equation} \label{attrrep::eq3}
f_T = L\frac{T-T_M}{T_M}  g(\phi),
\end{equation}
corresponding to a homogeneous overheating or undercooling with respect to the bulk melting temperature.
We use the simple choice
\begin{equation} \label{attrrep::eq5}
g(\phi) = 1- (\phi/\phi_0)^2(3+2\phi/\phi_0),
\end{equation}
which has the desired property that, for finite $\phi_0$, $g(\phi)$ varies from zero in the liquid to unity in the solid and has vanishing derivative in the bulk in order not to shift the equilibrium values of the phase fields. Notice that treating the case where $\phi_0$ is exactly zero would in principle require a coupling function that also depends on $\psi$. However such a complication is unecessary since we do not study this special case here, which was only discussed in the context of Fig. \ref{scattering} to motivate the model. The behavior of the model for $\phi_0=0$ and $\phi_0\ll 1$ are qualitatively very similar.

The total free-energy density is then given by $f=f_b+f_k +f_T$, and the free-energy $F$ is the volume integral of this expression. The stationary equations are
\begin{equation} \label{attrrep::eq4}
\frac{\delta F}{\delta \psi}=0, \qquad \frac{\delta F}{\delta \phi}=0.
\end{equation}

\subsection{Liquid film width and disjoining potential}

In the limit of an infinitely wide liquid layer, it is again sufficient to inspect the behavior in the vicinity of the liquid fixpoint $\psi=0, \phi=-\phi_0$, where the potential landscape becomes to second order
\begin{eqnarray}
f_b + f_T &=& \left[h(1+\phi_0^2)^2 + \frac{3}{\phi_0^2}L\frac{T-T_M}{T_M}  \right] (\phi+\phi_0)^2 \nonumber \\ 
&& + h(1+\phi_0^2)^2 \psi^2.  \label{attrrep:eq6}
\end{eqnarray}
Then, the linearized equations of ``motion'' are
\begin{eqnarray}
\sigma\alpha\ddot{\psi} &=& 2h(1+\phi_0^2)^2 \psi, \label{attrrep:eq7} \\
\sigma\ddot{\phi} &=& 2 \left[h(1+\phi_0^2)^2 + \frac{3}{\phi_0^2}L\frac{T-T_M}{T_M}  \right] (\phi+\phi_0), \label{attrrep:eq8}
\end{eqnarray}
with the general solution
\begin{eqnarray}
\psi(x) &=& c_{\psi+} \exp(\lambda_\psi x) + c_{\psi-} \exp(-\lambda_\psi x), \label{attrrep::eq9} \\
\phi(x) &=& c_{\phi+} \exp(\lambda_\phi x) + c_{\phi-} \exp(-\lambda_\phi x) -\phi_0, \label{attrrep::eq10}
\end{eqnarray}
and growth rates
\begin{eqnarray}
\lambda_\psi &=& \frac{\sqrt{2h}(1+\phi_0^2)}{\sqrt{\sigma\alpha}}, \label{attrrep::eq11} \\
\lambda_\phi &=& \sqrt{\frac{2h(1+\phi_0^2)^2}{\sigma}+\frac{6}{\phi_0^2}L\frac{T-T_M}{T_M\sigma}}. \label{attrrep::eq12} 
\end{eqnarray}
Symmetry of the solution according to Fig.~\ref{attrrep::fig2} demands $c_{\psi-}=-c_{\psi+}$ and $c_{\phi-}=c_{\phi+}$.
The Hamiltonian $H=f_k-f_b-f_T$ becomes in quadratic approximation by energy conservation
\begin{equation} \label{attrrep::eq13}
H= c_{\psi+}^2 \sigma\left[ 2\alpha\lambda_\psi^2 - 2\lambda_\phi^2 (c_{\phi+}/c_{\psi+})^2 \right]   = -L\frac{T-T_M}{T_M}.
\end{equation}
Notice that in the case $\alpha=1$ the two exponential decays become equal when we approach the melting point, and this case will thus require some additional care.
 Let us therefore discuss $\alpha\neq 1$ first.
 
\begin{figure}
\begin{center}
\includegraphics[width=8.5cm]{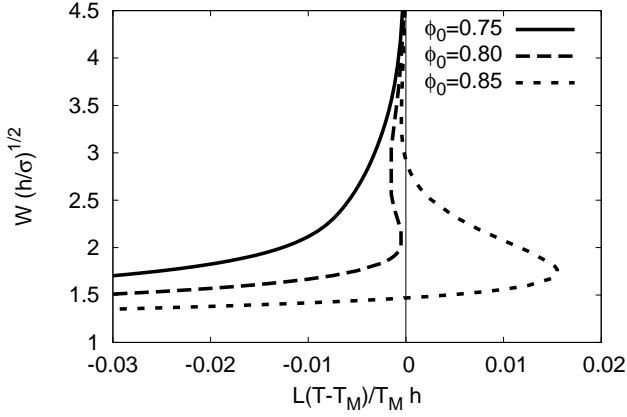}
\caption{Liquid film width as function of dimensionless overheating for $\alpha=1.25$. The parts of the curves with a negative slope are unstable. In this model, the interface thickness $\xi=(\sigma/h)^{1/2}$ and $\gamma_{sl}\sim (\sigma h)^{1/2}$ so that the barrier height of the double well-potential $h\sim \gamma_{sl}/\xi$.
}
\label{attrrep::fig3}
\end{center}
\end{figure}

 For $T\to T_M$ only the field with the slowest decaying exponential contributes, which is $\psi$ for $\alpha>1$ and $\phi$ for $\alpha<1$.
If the two solid-liquid interfaces are far away from each other, they look (almost) the same as two single solid-liquid interfaces, located at $\pm W/2$, where $W$ is the liquid layer thickness.
We have asympotically e.g.~for $-W/2\ll x\ll 0$: $\psi\simeq-\Psi \exp(-\lambda_\psi(x+W/2))$ and $\phi\simeq-\phi_0+ \Phi\exp[-\lambda_\phi(x+W/2)]$, where both coefficients $\Phi$ and $\Psi$ are of order unity.
Matching this to the above general solution (\ref{attrrep::eq9}, \ref{attrrep::eq10}) gives then 
\begin{eqnarray}
c_{\psi+} = -c_{\psi-} &=& \Psi\exp(-\lambda_\psi W/2), \label{attrrep::eq14}\\
c_{\phi+} = c_{\phi-} &=& \Phi\exp(-\lambda_\phi W/2). \label{attrrep::eq15}
\end{eqnarray}
In the limit $W\to\infty$ the weight factor $c$ in front of the faster decaying mode (larger $\lambda$) is exponentially suppressed in comparison to the other, and we can therefore drop its contribution in the Hamiltonian (\ref{attrrep::eq13}).
Hence, we can immediately conclude that the model is asymptotically repulsive for $\alpha>1$, because solutions can exist only for $T<T_M$, and vice versa for $\alpha<1$.
Notice that the value of $\phi_0$ is not relevant for this general long range interaction character;
the asymptotic analysis makes predictions only for the limit $T\to T_M$ (or equivalently $W\to\infty$), but the value $\phi_0$ can still significantly change the solutions with a liquid layer thickness of the order of the interface thickness.
This behavior is shown in Fig.~\ref{attrrep::fig3} for $\alpha=1.25$.
Obviously, all cases are repulsive at large distances, but nevertheless a larger value of $\phi_0$ changes the pure repulsion and introduces a short scale attraction (with stable and unstable solutions above the melting temperature) and a first order transition character.
Here, we defined $W$ as the distance between the points where $\psi$ crosses the values $-1/2$ and $1/2$ respectively.

\begin{figure}
\begin{center}
\includegraphics[width=8.5cm]{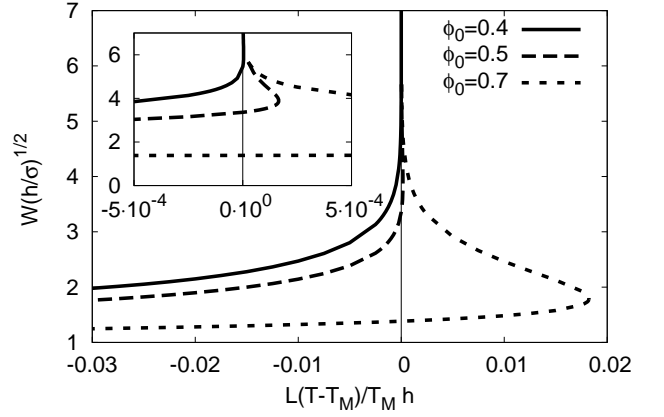}
\caption{Liquid film width as function of dimensionless overheating for $\alpha=0.75$. The parts of the curves with a negative slope are thermodynamically unstable.}
\label{attrrep::fig4}
\end{center}
\end{figure}

The corresponding long-distance attraction is depicted in Fig.~\ref{attrrep::fig4} for $\alpha=0.75$, although the behavior is here less pronounced.

To compare these results to analytical predictions, we need to determine $\Psi$ and $\Phi$ (they are functions of $\phi_0$ and $\alpha$) first;
since, in contrast to the multi-order parameter model of Ref. \cite{Plapp}, the profile of a single solid-liquid interface is not known analytically, we have to find the matching constants numerically as follows:
We set up a single solid-liquid interface at $T=T_M$, so the interface does not move.
Assuming that the interface, i.e. the point $\psi=1/2$, is located at $x=0$ and the solid phase in the domain $x<0$, we can look at the decay into the liquid region.
For $x\gg 0$, this decay is exponential, and we match it to $\psi\simeq \Psi \exp(-\lambda_\psi x)$ and $\phi\simeq \Phi \exp(-\lambda_\phi x) -\phi_0$, from which we get the desired prefactors $\Psi$ and $\Phi$.

We can then extract the asymptotic behavior of the liquid layer thickness and the disjoining potential as before.
From the energy balance (\ref{attrrep::eq13}) and the exponential prefactors (\ref{attrrep::eq14}, \ref{attrrep::eq15}) we get immediately
\begin{equation} \label{attrrep::eq16}
W \simeq \left\{ \begin{array}{cc} 
\displaystyle
-\frac{\alpha^{1/2}(\sigma/h)^{1/2}}{\sqrt{2}(1+\phi_0^2)} \ln \frac{-L(T-T_M)}{4\Psi^2 (1+\phi_0^2)^2 T_M h} & \mathrm{for }\; \alpha>1 \\
\displaystyle
-\frac{(\sigma/h)^{1/2}}{\sqrt{2}(1+\phi_0^2)} \ln \frac{L(T-T_M)}{4\Phi^2 (1+\phi_0^2)^2 T_M h} & \mathrm{for }\; \alpha<1 
\end{array} \right.
\end{equation}
The first asymptotic expression is of course applicable only for $T<T_M$, the second only above the melting point.
Notice that here we had to evaluate (\ref{attrrep::eq12}) at $T=T_M$ for the lowest order result.
From this and the asymptotic relation $V_{ex}'(W)\simeq L(T-T_M)/T_M$ we get for the disjoining potential
\begin{eqnarray}
V_{ex}(W) &\simeq& 
2\sqrt{2} \Psi^2 (1+\phi_0^2) \alpha^{1/2} (\sigma h)^{1/2} \times \nonumber \\
&& \times \exp \left[ -\frac{\sqrt{2}(1+\phi_0^2)}{\alpha^{1/2}(\sigma/h)^{1/2}} W \right]   \label{attrrep::eq16a}
\end{eqnarray}
for $\alpha>1$ and
\begin{eqnarray}
V_{ex}(W) &\simeq& -2\sqrt{2} \Phi^2 (1+\phi_0^2) (\sigma h)^{1/2} \times \nonumber \\
&& \times \exp \left[ -\sqrt{2} (1+\phi_0^2) (\sigma/h)^{-1/2} W \right]   \label{attrrep::eq16b}
\end{eqnarray}
for $\alpha <1$.

Let us now look at the marginal case $\alpha=1$.
There, at the melting point both exponentials have the same decay rate, and $\psi$ leads to repulsion, whereas $\phi$ gives rise to attraction (which follows readily from the expression of the Hamiltonian), and it depends on the prefactors which effect is stronger.
It will turn out, that the transition between attraction and repulsion is then controlled by $\phi_0$, in agreement with the mechanical interpretation that the scattering angle for small $\phi_0$ is small.

\begin{figure}
\begin{center}
\includegraphics[width=8.5cm]{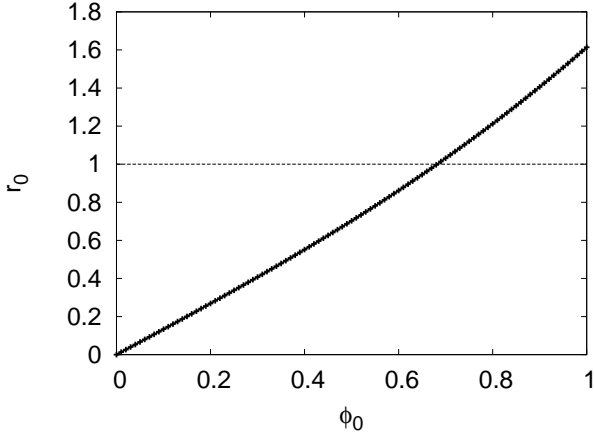}
\caption{The ratio of the exponential prefactors as function of the liquid fixpoint position.
For $\phi_0<0.68$ the model is repulsive, and attractive above this threshold.}
\label{attrrep::figx}
\end{center}
\end{figure}

\begin{figure}
\begin{center}
\includegraphics[width=8.5cm]{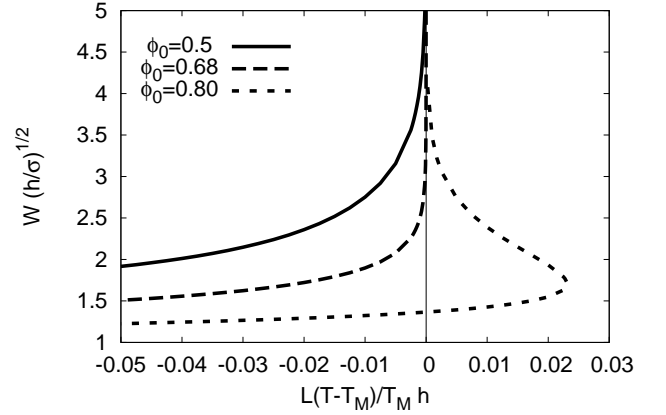}
\caption{Liquid film width as function of the dimensionless overheating, for $\alpha=1$.
For $\phi_0<0.68$ the model is repulsive, and attractive above this threshold.}
\label{attrrep::figy}
\end{center}
\end{figure}

We define the ratio of the exponential prefactors,
\begin{equation}
r := \frac{c_{\phi+}}{c_{\psi+}} \simeq \frac{\phi(x)+\phi_0}{\psi(x)} \exp[(\lambda_\psi-\lambda_\phi)x].
\end{equation}
Notice that this expression does not depend on temperature in the limit $T\to T_M$, i.e. $r=r_0 + {\cal O}[(T-T_M)/T_M]$.
We can therefore determine the constant $r_0$ numerically from an isolated solid-liquid interface at $T=T_M$, and the result is shown in Fig.~\ref{attrrep::figx}.
The reason why this is sufficient is that  in the expression for the Hamiltonian (\ref{attrrep::eq13}) the common prefactor $c_{\psi+}^2$ is already of order $(T-T_M)/T_M$ (as the right hand side), and therefore we need to evaluate the expression in brackets only at the melting temperature, i.e.~to the order $[(T-T_M)/T_M]^0$ (exactly at the melting temperature the liquid layer is infinitely wide and therefore the exponential prefactors are zero).
Then we get the solvability condition
\begin{equation}
L\frac{T-T_M}{T_M}= -4 h (1+\phi_0^2)^2 c_{\psi+}^2(1-r_0^2).
\end{equation}
Obviously, this equation has asymptotic solutions below the melting temperature only if $r_0^2<1$, which is the case for $|\phi_0| < 0.68$, and then the model is repulsive at large distances.
The numerical results confirm this prediction, see in Fig.~\ref{attrrep::figy}.

\begin{figure}
\begin{center}
\includegraphics[width=8.5cm]{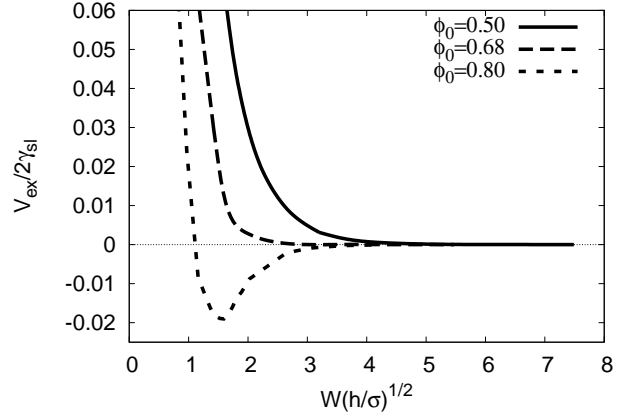}
\caption{Disjoining potentials versus dimensionless film width for $\alpha=1$ and three different values of $\phi_0$.}
\label{attrrep::fig5}
\end{center}
\end{figure}

We can again calculate the asymptotic behavior analytically, and obtain for the liquid layer thickness
\begin{equation} \label{attrrep::eq17}
W\simeq -\frac{(\sigma/h)^{1/2}}{\sqrt{2}(1+\phi_0^2)} \ln \frac{-L(T-T_M)}{4 (1+\phi_0^2)^2 \Psi^2 (1-r_0^2)T_M h}.
\end{equation}
Similarly, for the disjoining potential
\begin{equation} \label{attrrep::eq18}
V_{ex}(W) \simeq 2\Psi^2 \lambda_\psi(1-r_0^2) (\sigma h)^{1/2} \exp(-\lambda_\psi W)
\end{equation}
for $W\to\infty$.
The full disjoining potential, as obtained from the numerical simulation, is shown in Fig.~\ref{attrrep::fig5}.
For the chosen parameters, the model is repulsive at short distances even for a long range attraction (hard core repulsion).

\begin{figure}
\begin{center}
\includegraphics[width=8.5cm]{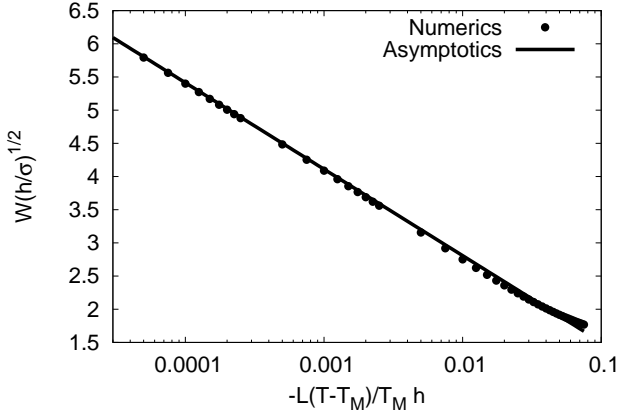}
\caption{Semi-logarithmic plot of the liquid layer thickness as function of temperature (below the melting point). The data is compared to the analytical prediction Eq.~(\ref{attrrep::eq17}). Parameters are $\alpha=1$ and $\phi_0=0.5$ here.
}
\label{attrrep::fig6}
\end{center}
\end{figure}

\begin{figure}
\begin{center}
\includegraphics[width=8.5cm]{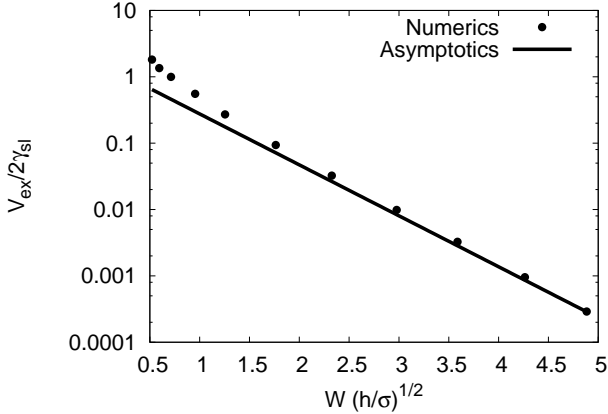}
\caption{Semi-logarithmic plot of the disjoining potential as function of the liquid layer thickness $W$. The data is compared to the analytical prediction Eq.~(\ref{attrrep::eq18}). We use $\alpha=1$ and $\phi_0=0.5$ here.
}
\label{attrrep::fig7}
\end{center}
\end{figure}

The analytical predictions are compared to the numerical results in Fig.~\ref{attrrep::fig6} for the liquid layer thickness and the disjoining potential in Fig.~\ref{attrrep::fig7} for $\alpha=1$ and $\phi_0=0.5$ (repulsive), which confirms the analysis.

We therefore conclude, that the proposed model can describe both long-range attraction and repulsion.
Despite its simplicity the parameters can be tuned to capture generic effects of many relevant materials.
For $\alpha>1$, the model can also display bistability (coexistence) of ``dry" and ``wet'' grain boundary states with different widths as shown in Fig. \ref{attrrep::fig3} for an intermediate value of $\phi_0$. This bistability has also been predicted by a frame-invariant phase-field model of a bicrystal \cite{LobWar02}. However, it has so far not been observed in molecular dynamics simulations of pure materials \cite{Hoyt09,NIST2,Jinetal09}. As shown in the next section, we find that solute addition can lead to bistability even for parameters of the model where bistability is absent in the pure limit.

\section{Solute effects}
\label{alloy}

We now extend the model to dilute alloys, corresponding to a phase diagram with straight solidus and liquidus lines. This dilute limit is described by adding to the free-energy density the contribution due to solute addition
\begin{equation} \label{alloy::eq1}
f_c = \frac{R T_M}{v_0} (c\ln c - c) + \chi(\phi) \Delta\epsilon\, c,
\end{equation}
where $R$ is the gas constant, $v_0$ is the molar volume, and $c$ is the mole fraction of solute
assumed much smaller than unity. This contribution includes the standard entropy of mixing term and a partitioning term that distinguishes between the energy density of impurities in solid and liquid via the coupling function $\chi(\phi)$.  
This function varies from $0$ in the liquid to $1$ in the solid and may be chosen equal to $g(\phi)$.
A dependence on $\psi$ can also be introduced to influence the segregation of impurities at the grain boundary, but we do not investigate this effect here.

The concentration field obeys in equilibrium the condition
\begin{equation} \label{alloy::eq2}
\frac{\delta F}{\delta c}=\mu.
\end{equation}
Since it enters the free-energy functional without gradient terms (in the mechanical analog, the ``coordinate'' $c$ belongs to a particle without mass, which follows the motion of the phase fields ``instantaneously''), we can eliminate it and rewrite the phase field equations as derived from the grand potential.
We obtain from the expression above
\begin{equation} \label{alloy::eq3}
c(\phi) = c_l^{(eq)} \exp\left[ -\frac{v_0\Delta\epsilon}{R T_M} \chi(\phi) \right]
\end{equation}
where we defined
\begin{equation} \label{alloy::eq4}
c_l^{(eq)} := \exp\left( \frac{v_0\mu}{R T_M} \right).
\end{equation}
Here, we immediately identify the meaning of the partition coefficient $k$, 
\begin{equation}
k = \exp\left( -\frac{v_0\Delta\epsilon}{R T_M} \right),
\end{equation}
since we get for the concentrations of an (infinite) solid-liquid equilibrium system 
\begin{equation} \label{alloy::eq4a}
c_s^{(eq)}=k c_l^{(eq)}.
\end{equation}
Notice that for a thin liquid layer the concentration differs there from the expression (\ref{alloy::eq4}), since the phase field $\phi$ does not fully reach the liquid value $-\phi_0$.

\begin{figure}
\begin{center}
\includegraphics[width=8.5cm]{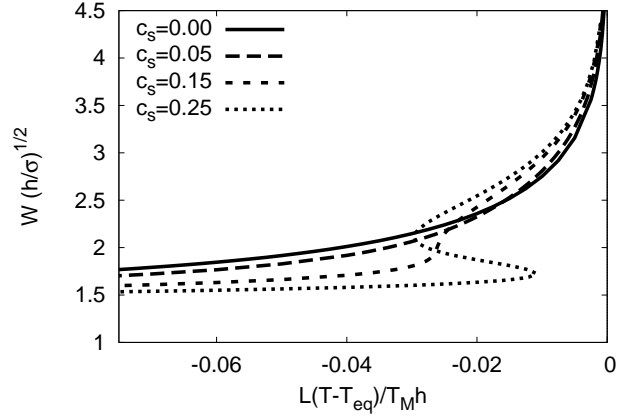}
\caption{Liquid film width as function of overheating for different solid concentrations. The addition of impurities makes the model more repulsive at large distances, and leads to a first order character at shorter liquid layer thicknesses. Parameters are $\alpha=1.0$, $\phi_0=0.5$, $m L/T_M h=-1$, $k=0.5$.}
\label{alloy::fig1}
\end{center}
\end{figure}

\begin{figure}
\begin{center}
\includegraphics[width=8.5cm]{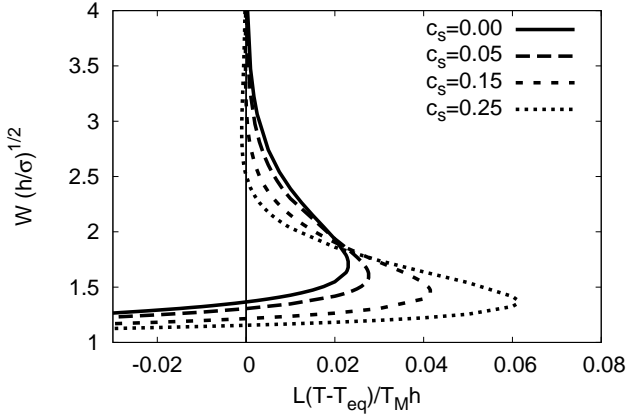}
\caption{Liquid film width as a function of dimensionless overheating.  Whereas the pure case is attractive, the addition of impurities leads to a long-range repulsion. Parameters are $\alpha=1.0$, $\phi_0=0.8$, $m L/T_M h=-1$, $k=0.5$.}
\label{alloy::fig2}
\end{center}
\end{figure}

We can change the ensemble and eliminate the conserved field and replace it by the intensive variable $\mu$.
The adequate thermodynamic functional is then the grand potential, from which the phase field equations can be derived variationally.
This implies that the mechanical analog holds with the potential energy of the particle now determined by the grand potential density instead of the free-energy density.
Then, the impurity contribution to the grand potential, $\omega_c=f_c-\mu c$, is
\begin{equation}
\omega_c = -\frac{R T_M}{v_0} c(\phi).
\end{equation}
Again, for the equilibrium of two bulk phases, the grand potential
\begin{equation}
\Omega(\psi, \phi, \mu) = \int \omega\, dV
\end{equation}
must be minimized, i.e.
\begin{equation}
\frac{\delta\Omega}{\delta\psi}=0,\qquad \frac{\delta\Omega}{\delta\phi}=0,
\end{equation}
which implies that its density is equal in solid and liquid for $W\to\infty$.
Here, $\omega = \omega_c+f_b+f_T+f_k$.
Since $f_b=f_k=0$ in both infinitely large bulk states, we get immediately $\omega_c(\phi=0) + L(T_{eq}-T_M)/T_M = \omega_c(\phi=-\phi_0)$, and therefore 
\begin{equation} \label{alloy::eq4b}
m c_l^{(eq)} = T_{eq} - T_M,
\end{equation}
which describes the straight liquidus line with slope
\begin{equation}
m = -\frac{RT_M^2}{v_0 L} (1-k).
\end{equation}
Expanding again up to second order around the liquid fixpoint we get
\begin{equation}
\omega_c = \frac{mL}{T_M(1-k)} c_l^{(eq)} + \frac{1}{2} \frac{mL}{T_M(1-k)} \chi''\, c_l^{(eq)} (\ln k) (\phi+\phi_0)^2 
\end{equation}
with $\chi'' = \chi''(\phi=-\phi_0)$.
From the total grand potential we get the linearized equations of motion with $\omega_p=\omega-f_k$
\begin{eqnarray}
\sigma\ddot{\phi} &=& \frac{\partial\omega_p}{\partial\phi} = \sigma \lambda_{\phi, i}^2 (\phi+\phi_0), \\
\sigma \alpha\ddot{\psi} &=& \frac{\partial \omega_p}{\partial\psi} = \sigma \alpha \lambda_\psi^2 \psi, 
\end{eqnarray}
where we defined
\begin{eqnarray}
\lambda_{\phi, i} &=& \Bigg( \frac{2h(1+\phi_0^2)^2}{\sigma} + \frac{6}{\phi_0^2} L \frac{T_{eq}-T_M}{T_M\sigma} \nonumber \\
&& + \frac{mL}{T_M (1-k)\sigma} c_l^{(eq)} (\ln k) \chi'' \Bigg)^{1/2}.
\end{eqnarray}
The solution for the linearized phase fields has again the structure (\ref{attrrep::eq9}, \ref{attrrep::eq10}).
Obviously, the decay rate of the liquid field $\phi$ is modified in comparison to the pure case, and it becomes larger here.
This means that the model becomes more repulsive through alloying, and the effect is more pronounced for stronger partitioning.
For the particular choice $\chi=g$, with $g$ being the thermal coupling function (\ref{attrrep::eq5}) we obtain
\begin{equation}
\lambda_{\phi, i} = \left[ \frac{2h(1+\phi_0^2)^2}{\sigma} + \frac{6}{\phi_0^2\sigma} L \frac{T_{eq}-T_M}{T_M} \left(1+\frac{\ln k}{1-k} \right) \right]^{1/2},
\end{equation}
which has to be compared to the decay rate of $\psi$ given by Eq.~(\ref{attrrep::eq11}).
The influence of impurities is shown in Figs.~\ref{alloy::fig1} and \ref{alloy::fig2} as function of the temperature deviation from equilibrium and fixed chemical potential;
for all numerical calculation $\chi=g$ is used.
For convenience, the latter quantity is expressed through the concentration in the solid far away from the interfaces.
It is equivalent to the notion of the chemical potential through the relations (\ref{alloy::eq4}) and (\ref{alloy::eq4a}).
In both cases, the addition of impurities leads to a pronounced first order character, and an enhanced repulsion at large distances with higher impurity concentration.
For the particular marginal choice $\alpha=1$ in Fig.~\ref{alloy::fig2}, which is attractive in the pure case by the choice of $\phi_0$, the model becomes immediately repulsive through the presence of impurities.

\begin{figure}
\begin{center}
\includegraphics[width=6cm]{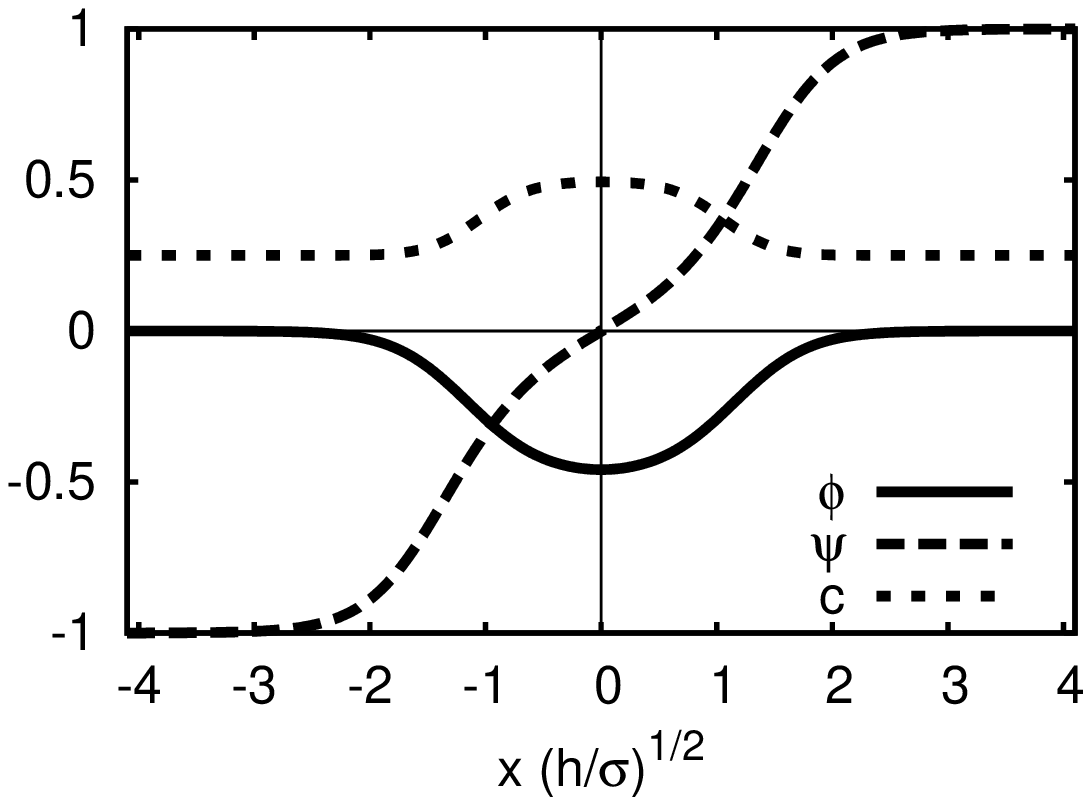}
\includegraphics[width=6cm]{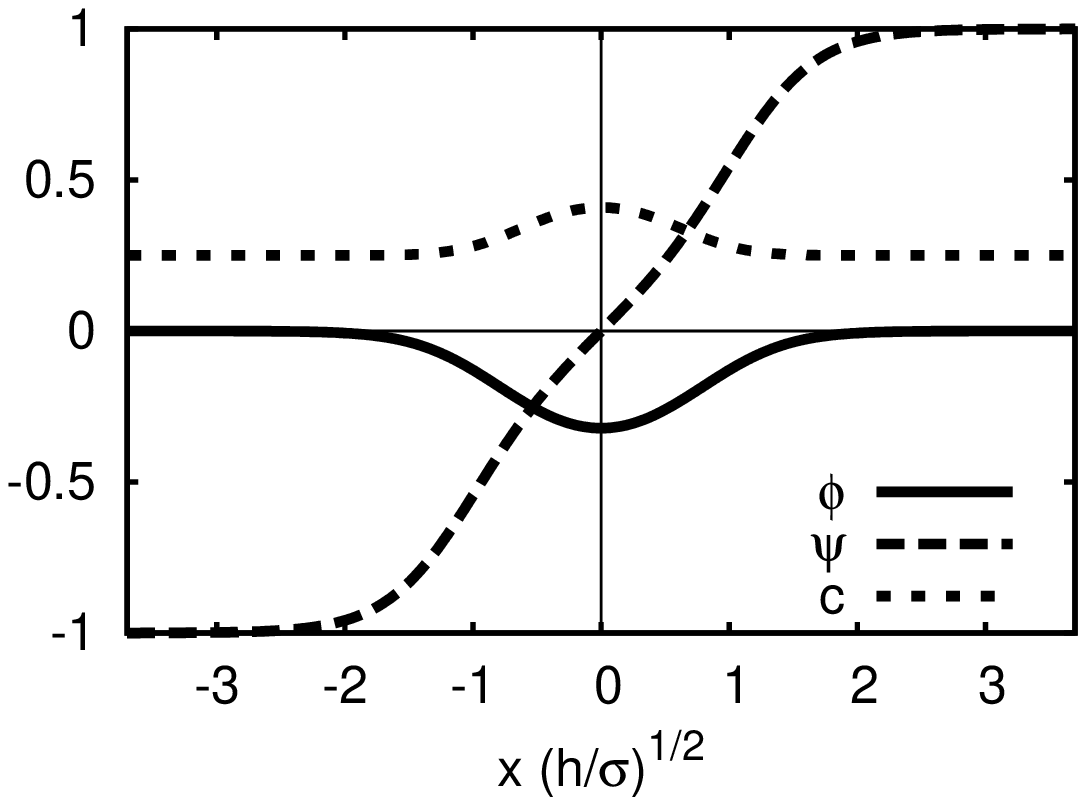}
\includegraphics[width=6cm]{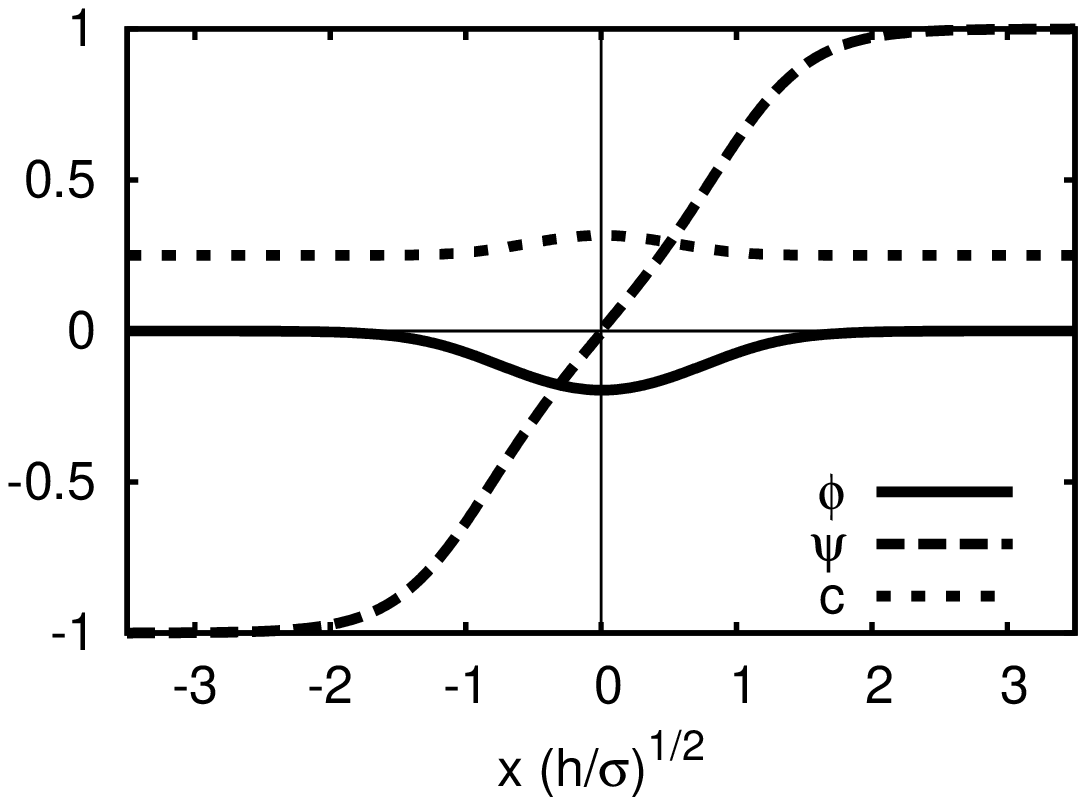}
\caption{Profiles of phase fields and impurity concentration for a temperature inside the bistable range of Fig.~\ref{alloy::fig1}; $c_s=0.25$ and $L(T-T_{eq})/T_Mh=-0.02$. For this temperature, three solutions exist with $W$ decreasing from top to bottom. The top and bottom solutions can coexist whereas the middle solution is thermodynamically unstable.}
\label{alloy::concprofiles}
\end{center}
\end{figure}

Fig.~\ref{alloy::concprofiles} shows the profiles of the phase fields $\phi$ and $\psi$ and the concentration $c$ as function of the position for parameters as in Fig.~\ref{alloy::fig1}, $c_s=0.25$ and $L(T-T_{eq})/T_Mh=-0.02$.
Here, two stable and one unstable solution exists, which differ by the melt layer thickness.
For the solution with the widest liquid layer a rather pronounced liquid phase exists, i.e.~the phase field $\phi$ is almost stationary in the center, but it does not fully reach its bulk equilibrium value $-\phi_0$.
Consequently, also the impurity concentration is significantly larger than in the bulk solid.
For the solutions with the thinner width, the grain boundary is almost dry, and the concentration only slightly increased.

\begin{figure}
\begin{center}
\includegraphics[width=8.5cm]{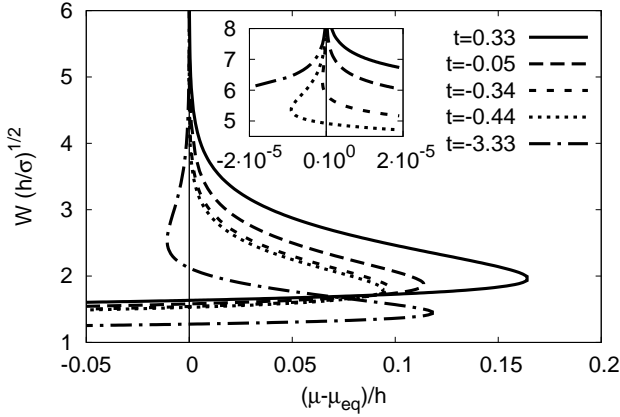}
\caption{Liquid film width as function of the chemical potential deviation from bulk equilibrium for different values of the dimensionless reduced temperature $t$ (see text).
The inset shows a magnification around the origin to demonstrate the transition from an attractive to a repulsive behavior if the temperature is lowered.
Parameters are $\alpha=0.9$, $\phi_0=0.7$, $m L/T_M h=-1$ and $k=0.5$.}
\label{alloy::fig3}
\end{center}
\end{figure}

\begin{figure}
\begin{center}
\includegraphics[width=8.5cm]{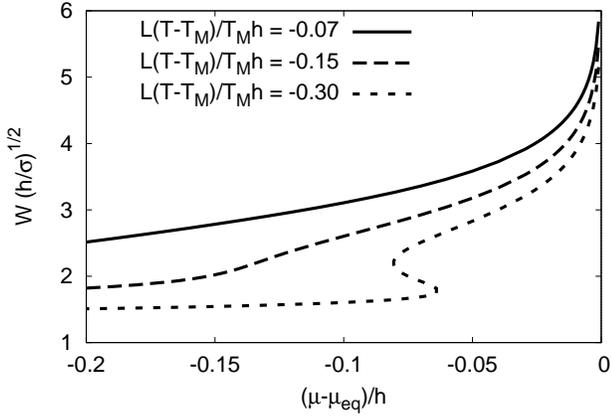}
\caption{Liquid film width as function of the chemical potential deviation from bulk equilibrium for different values of the dimensionless overheating. Parameters are $\alpha=1.25$, $\phi_0=0.75$, $m L/T_M h=-1$ and $k=0.5$.}
\label{alloy::fig4}
\end{center}
\end{figure}

Again, for $\chi=g$, starting with an attractive situation with $\alpha<1$ without impurities, the long-range interaction becomes repulsive for
\begin{equation}
h(1+\phi_0^2)^2 \left( \frac{1}{\alpha}-1 \right) = \frac{3 L}{\phi_0^2} \frac{T_{eq}-T_M}{T_M} \left( 1+\frac{\ln k}{1-k} \right),
\end{equation}
because then the decay lengths of the ``attractive'' field $\phi$ and the ``repulsive'' field $\psi$ become equal at the coexistence point;
the solution of this equation defines a critical temperature $T^*(\alpha, \phi_0, m, k)$.
We can then define a (dimensionless) deviation from this value as $t=(T^*-T)/T^*$, and the numerical results are shown in Figs.~\ref{alloy::fig3} and \ref{alloy::fig4}.
Here, we keep the temperature constant and vary the chemical potential, and the behavior is qualitatively similar to the curves with fixed chemical potential and varying temperature.
The equilibrium chemical potential $\mu_{eq}$ is given by the expressions (\ref{alloy::eq4}), (\ref{alloy::eq4a}) and (\ref{alloy::eq4b}).
Here, from the given temperature $T$ the equilibrium chemical potential $\mu_{eq}$ can be calculated;
a change of the ``supplied'' composition in the solid phase far away from the grain boundary allows then to vary the chemical potential.
The results confirm the analytical prediction that the system becomes repulsive at long distances below the critical temperature $T^*$.

\begin{figure}
\begin{center}
\includegraphics[width=8.5cm]{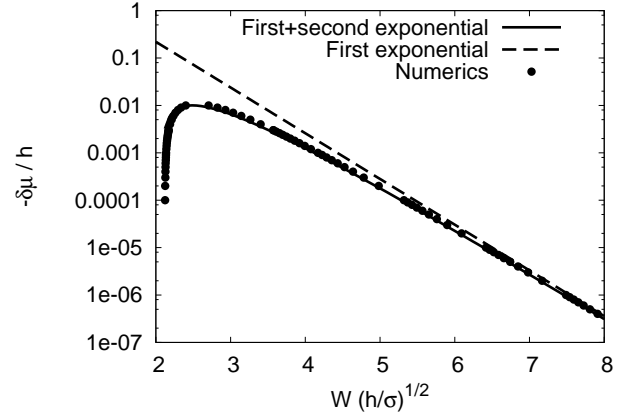}
\caption{Comparison of the numerical data for the deviation of the chemical potential $\delta\mu$ versus the melt layer thickness $W$ with the asymptotic prediction for a binary alloy. The parameters are $\alpha=0.9$, $\phi_0=0.7$, $k=0.5$, $m L/T_M h=-1$ and $L(T-T_M)/T_Mh=-0.45$. At large distances the interfaces repel each other, and the contribution of the second exponential leads to an attraction. Asymptotically, the behavior is determined by the slowest decaying exponential only; for smaller separations, the behavior is well described by the contribution from the two slowest decaying modes.}
\label{alloy::fig5}
\end{center}
\end{figure}

In the same way as before, we can calculate the asymptotic energy balance using the fact that the
Hamiltonian $H=f_k-\omega_p$ is constant, which yields
\begin{equation}
-c_l^{(eq)}(1-k) \delta\mu = 2c_{\psi+}^2\alpha \sigma \lambda_\psi^2 - 2c_{\phi+}^2\sigma \lambda_{\phi, i}^2,
\end{equation}
with $\delta\mu =  \mu-\mu_{eq}$.
The logarithmic divergence of the liquid layer thickness at $\mu_{eq}$ follows again immediately from the preceding relations, and we can take into account also the effect of the second exponential, resulting in the implicit relation
\begin{eqnarray}
\delta\mu &=& - \frac{1}{c_l^{(eq)}(1-k)} \Big( 2\sigma \Psi^2 \exp(-\lambda_\psi W) \alpha \lambda_\psi^2 \nonumber \\
&& - 2\sigma \Phi^2 \exp(-\lambda_{\phi, i} W) \lambda_{\phi, i}^2 \Big),
\end{eqnarray}
where the matching constants $\Psi(\alpha, \phi_0, m, k, T)$ and $\Phi(\alpha, \phi_0, m, k, T)$ are determined from a single solid-liquid interface at bulk equilibrium as before.
For large separation, of course only the slowest decaying exponential contributes, but the inclusion of the next term can lead to a substantial better agreement with the numerically obtained result, as shown in Fig. \ref{alloy::fig5}.

\begin{figure}
\begin{center}
\includegraphics[width=8.5cm]{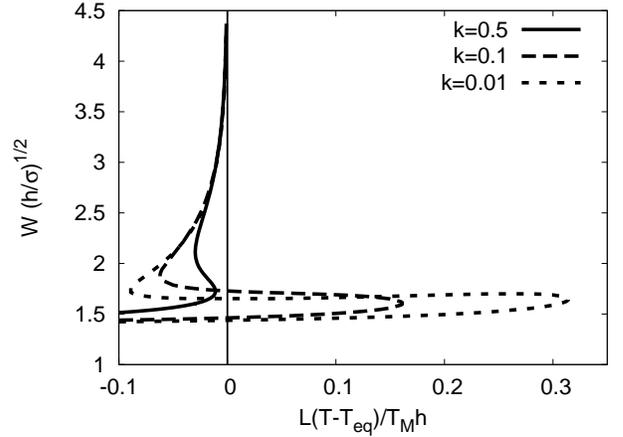}
\caption{Liquid film width as function of overheating for a repulsive case and different partition coefficients. 
The chemical potential is the same in all cases, expressed through the equilibrium liquid concentration $c_l^{(eq)}=c_s^{(eq)}/k=0.5$. The other parameters are $\alpha=1$, $\phi_0=0.5$ and $mL/T_M h=-1$.}
\label{alloy::fig6}
\end{center}
\end{figure}

Finally, we checked the influence of the partition coefficient on the results.
Fig.~\ref{alloy::fig6} shows the melt layer thickness as function of temperature for fixed chemical potential.
We see in general that a stronger partitioning system can be overheated more.
In particular, for the example shown here, no equilibrium solution exists above the melting point for $k=0.5$, whereas alloys with smaller partition coefficient exhibit stable and unstable solutions also above the melting point.

We have only studied here a free-energy functional without a gradient term  
$\sim(\nabla c)^2$. With the inclusion of such as term,
the equilibrium condition (\ref{alloy::eq2}) has the same form as the phase field equation (\ref{attrrep::eq4}). Therefore, the same type of analysis that exploits a mechanical analog can be performed with a concentration field that now possesses a ``mass''.

In summary, 
a pure system that is attractive at melting, can become repulsive with solute addition above some threshold concentration. Furthermore, solute addition can lead to bistability (i.e., existence of stable and metastable states at the same temperature on either side of a Maxwell point) with the effect being more pronounced for stronger partitioning (smaller $k$). A numerical estimate of the temperature range of bistability is useful to examine if first order hysteretic transitions between liquid films of different widths could be observed. In Fig. \ref{alloy::fig6}, bistability extends over a dimensionless temperature range $|L(T-T_{eq})/(T_Mh)|$ of almost $0.1$ for $k=0.01$. Since $h\sim \gamma_{sl}/\xi$, we obtain that $|T-T_{eq}|/T_M\sim 0.1\gamma_{sl}/(L\xi)$. With $\xi\sim 1$ nm and typical values of $L$ and $\gamma_{sl}$ for metallic systems (e.g., $L\approx  3\times 10^9$ J/m$^3$ and $\gamma_{sl}\approx  0.3$ J/m$^2$ for pure Ni), we
obtain that $|T-T_{eq}|/T_M\sim 10^{-2}$. So, according to the present model, bistability should be present over a temperature range below melting of the order of tens of degrees.

\section{Stress effects}
\label{elasticity}

Stress effects can have a strong influence on microstructural evolution in the presence of dry and wet grain boundaries at high homologous temperature, as in the practical case of hot cracking of metallic alloys \cite{Rappaz03}.
Here we limit our study to the effect of a uniaxial stress applied along an axis perpendicular to the grain boundary. We couple phase change and elasticity by modeling the liquid as a solid with vanishing shear modulus, as in previous studies of solidification under stress \cite{Slutsker}, the Asaro-Tiller-Grinfeld instability \cite{Kassner},  and fracture associated with a phase change \cite{Spatschek}.
In this approach, the local displacement vector (with two degrees of freedom in two dimensions) is also represented in the liquid, whereas the local state in this phase is fully characterized by the hydrostatic pressure.
The remaining degree of freedom allows to represent the slip of a liquid at the solid-liquid interface, as an incoherent boundary between two solids.
For the present purpose of a one-dimensional analysis, the situation is even simpler since we assume that the liquid film cannot expand in the direction parallel to the interface.
Then solidification shrinkage due to a density difference between solid and liquid can be rigorously described as an ``eigenstrain''.

Since the focus is here on interactions on short scales, the issue of proper coupling of the elastic fields to the local phase field arises.
As before for the temperature and impurity coupling, this relationship is not unique, and different choices can lead to the same sharp interface limit.
Therefore, additional physical assumptions or input from other sources is required.
We discuss here two different choices of the coupling function to illustrate the consequences of this effect.

The first choice for the additional free-energy contribution is given by
\begin{equation}
f_{el} = \frac{1}{2} \bar{\lambda}\bar{\epsilon}_{jj}^2 + \bar{\mu} \bar{\epsilon}_{jk}^2
\end{equation}
with 
\begin{equation}
\bar{\epsilon}_{ij} = \epsilon_{ij} - [1-g(\phi)] \epsilon_{ij}^{(0)}.
\end{equation}
The strain is derived from the displacements $u_i$ as  
\begin{equation}
\epsilon_{ij} = \frac{1}{2} \left(\frac{\partial u_i}{\partial x_j} + \frac{\partial u_j}{\partial x_i} \right).
\end{equation}
Here, we assign the eigenstrain $\epsilon_{ij}^{(0)}$ to the liquid phase, and we use the same coupling function $g(\phi)$ as above.
Also, we define averaged elastic constants
\begin{equation}
\bar{\lambda} = g(\phi) \lambda_s + [1-g(\phi)] \lambda_l,
\end{equation}
where $\lambda_s$ and $\lambda_l$ are the first Lam\'{e} coefficients of the solid and liquid phase, respectively.
Similar definitions are used for the shear modulus $\bar{\mu}$.
The entire free-energy $F$ depends then also additionally on the displacement field, and equilibrium requires that $F$ is also minimized with respect to this new degree of freedom, which implies static elasticity, $\partial\sigma_{ij}/\partial x_j=0$.
The stress tensor is here given by Hooke's law,
\begin{equation}
\sigma_{ij} = 2\bar{\mu} \bar{\epsilon}_{ij} + \bar{\lambda} \bar{\epsilon}_{kk} \delta_{ij}.
\end{equation}

The advantage of this model is that in the case of equal elastic constants in a one-dimensional case the influence of stress can be mapped to a temperature tilt:
We assume $\bar{\mu}=0$ (since the liquid phase has no elastic response to shear and we demand the equality of the elastic constants in both phases), $\lambda_s=\lambda_l\equiv\lambda$, 
and the only nonvanishing displacement component $u_x$ depends only on $x$.
Then elastic equilibrium,  $\delta F/\delta u_i=0$, requires that the stress $\sigma_{xx}$ is spatially constant, with
\begin{equation} \label{elast::eq1}
\sigma_{xx} = \lambda_s [\epsilon_{xx} - (1-g(\phi)) \epsilon_{ll}^{(0)} ].
\end{equation}
The elastic free-energy density becomes then
\begin{equation}
f_{el} = \frac{1}{2} \sigma_{xx}^2/\lambda_s.
\end{equation}
Notice that the proper underlying boundary conditions for a minimization of the total {\em free}-energy are fixed displacements;
although the stress is spatially constant it varies if the interfaces move, and we get the additional contribution to the equations of motion
\begin{equation}
\frac{\partial f_{el}}{\partial \phi} = \sigma_{xx} g'(\phi)\epsilon_{ll}^{(0)},
\end{equation}
which has exactly the same structure as the driving force term arising from the thermal tilt (\ref{attrrep::eq3}).
We can then immediately identify the driving force term in the phase field equation due to elasticity with the one which stems from thermal effects if we relate
\begin{equation}
L\frac{T-T_M}{T_M} \equiv \epsilon_{ll}^{(0)}\sigma_{xx},
\end{equation}
which is the classical Clausius-Clapeyron relation since $\epsilon_{ll}$ is the relative volume change , i.e. $\epsilon_{ll}^{(0)}=\Delta v/v$.
As a result, the application of a stress is equivalent to a change of temperature, and the abscissae of all previous plots of liquid film width versus dimensionless overheating in section \ref{attrrep} can be relabeled with $(\Delta v/v)\sigma_{xx}/h$ in place of $L(T-T_M)/(T_Mh)$. For an applied stress to change the liquid film width appreciably, it should produce an equivalent temperature change of at least a few degrees. For a typical relative volume change of a few percent and $L$  a few times $10^9$ J/m$^3$, this stress
$\sigma_{xx}=L(\Delta v/v)^{-1}(T-T_M)/T_M$ should be in the range of hundreds of MPa, and therefore of the same order as a typical yield stress.

Apart from situations with given {\em stress}, we can also consider the case of given {\em displacement}, and we can obtain it directly from the results for fixed stress.
From Eq.~(\ref{elast::eq1}) we obtain the total displacement
\begin{equation}
\delta := \int\limits_0^L \epsilon_{xx}\, dx = \frac{\sigma_{xx}}{\lambda_s} L + \epsilon_{ll}^{(0)} \int\limits_0^L [1-g(\phi)]dx,
\end{equation}
which depends now explicitly on the system size $L$.
We note that here always a ``macroscopic'' stress free solution exists with $\epsilon_{xx}=0$ in the solid and $\epsilon_{xx}=\epsilon_{ll}^{(0)}$ in the liquid (at $T=T_M$).
This implies
\begin{equation}
W\epsilon_{ll}^{(0)} = \delta
\end{equation}
independent of the elastic parameters;
this equation reflects that the applied displacement is compensated by the volume change during melting, and the liquid layer thickness adjusts itself such that the elastic energy is minimized (at $T=T_M$), i.e.~the system is stress free.
The above relation is of course a sharp interface prediction, therefore valid asymptotically for large $\delta/\epsilon_{ll}^{(0)}$.

\begin{figure}
\begin{center}
\includegraphics[width=8.5cm]{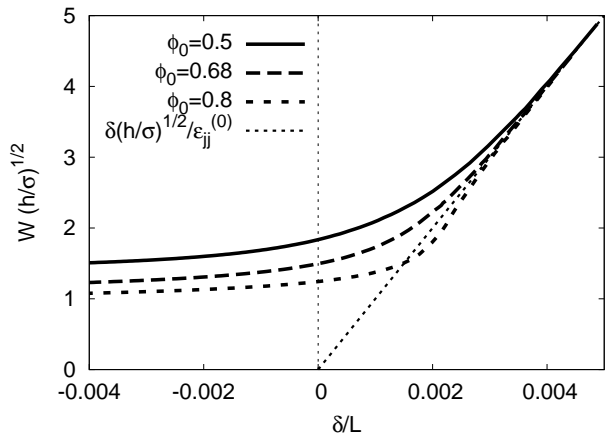}
\caption{Liquid film width as function of the imposed fixed displacement $\delta$ at solid surfaces at melting ($T=T_M$).
Parameters are $\alpha=1$, $L=100\,(\sigma/h)^{1/2}$, $\epsilon_{ll}^{(0)}=0.1$ and $\lambda/h=500$.}
\label{elast::fig1}
\end{center}
\end{figure}

The numerical results are shown in Fig.~\ref{elast::fig1}, and they exhibit the correct asymptotic behavior, which does not depend on the interaction of the interfaces, as it becomes negligible for large interface separations.
Notice that equilibrium liquid layer thicknesses for different system sizes look different if plotted versus the average {\em strain} $\delta/L$, but they collapse to the same curve if drawn as function of {\em stress}, provided that the system size is much bigger than the interface thickness, $L\gg (\sigma/h)^{1/2}$.
Also, it is worthwhile to mention that the stability of the branches has changed in comparison to a case with prescribed stress:
Whereas for fixed stress the asymptotic branch with negative slope is unstable, solutions on the macroscopic branch for fixed displacement are stable, since it corresponds to an energy minimum;
nevertheless they correspond to the same solution.
The same behavior will of course also occur for the pure thermal coupling if instead of the temperature the heat content is kept fixed.

We note that the stress-temperature duality is a consequence of the chosen coupling function and the proper free-energy functional and not necessarily true for other choices, and it holds in general only for equal elastic constants.
To make this effect more transparent, let us consider the special situation of a stress free state, where we get for the above coupling $f_{el}=0$.
An alternative way to implement the elastic energy is \cite{bms}
\begin{equation}
\tilde{f}_{el} = \frac{1}{2}\lambda_s \left( \epsilon_{jj}^2 g(\phi) + [1-g(\phi)] (\epsilon_{jj}-\epsilon_{jj}^{(0)})^2 \right),
\end{equation}
where we, for simplicity, directly assumed $\mu=0$ and equal Lam\'e coefficients.
It differs from the above expression only by the averaging in the interface region and has the same sharp interface behavior.
Now we get from the mechanical equilibrium condition,
\begin{equation} \label{elast::eq2}
\tilde{\sigma}_{xx}= \lambda_s\left( \epsilon_{xx} g(\phi) + (1-g(\phi)) (\epsilon_{xx} - \epsilon_{ll}^{(0)}) \right) = 0,
\end{equation}
where we assumed again a stress free situation.
Multiplication with $(\epsilon_{xx}-\epsilon_{ll}^{(0)})$ and integration over $x$ gives
\begin{equation}
0 = F_{el} - \frac{1}{2} \lambda_s \epsilon_{ll}^{(0)} \int\limits_0^L \epsilon_{xx}\, g(\phi)\, dx
\end{equation}
with $F_{el} = \int \tilde{f}_{el} dx$, which does not vanish in a stress free situation.
Notice that according to Eq.~(\ref{elast::eq2}) the integrand is zero in the bulk phases, therefore the additional integral term only renormalizes the interfacial energy, and this effect vanishes in the sharp interface limit.
Nevertheless, the above expression shows that in a stress free situation the elastic energy does not vanish in the interfacial region, and therefore a mapping of stress to temperature is no longer possible as before.

However, it is intuitively clear that in the sharp interface limit, where the choice of the interpolation becomes irrelevant, we always recover the Clausius-Clapeyron relation for the considered case of a diagonal eigenstrain, as expected.
In particular, the stress free branch for the case of fixed displacements is indeed a sharp interface result and does not depend on the coupling for $W\gg (\sigma/h)^{1/2}$.

\section{Concluding remarks}

In summary, we have developed an analytical approach to compute short-range forces between diffuse interfaces in multi-phase-field models based on a mechanical analog. Even though we have discussed our results in the context of grain boundary wetting, the approach is general and should be applicable to a broad class of problems. We have found that    multi-phase-field formulations generally considered in the literature do not describe repulsive interactions at large distance when the phase fields strictly represent local phase fractions. Motivated by this limitation, we have introduced a simple two-phase-field model with tunable interaction, which can represent both attractive and repulsive interactions.

This model was only developed here for a bicrystal. Therefore, it would be interesting to extend this formulation to represent an arbitrary number of grains of different crystal orientations, while retaining the flexibility to represent both attractive and repulsive interactions between different grains. Such a formulation should prove valuable to investigate strain localization in the context of hot cracking with coupling to elasticity and a shear modulus dependent on liquid film width.

Our study of solute and stress effects has yielded some interesting insights that warrant further investigations.

Firstly, above some threshold concentration, solute addition induces coexistence of different grain boundary states with different liquid film widths over a finite temperature range just below melting. This range is generally small but is estimated here to increase up to tens of degrees for a partition coefficient $k\sim 0.01-0.1$. Interestingly, the strength of this effect can be understood analytically to scale $\sim -\ln k$ in the dilute binary alloy limit. To our knowledge, this type of bistability has not yet been clearly observed in molecular dynamics simulations or experiments, although its existence has been predicted in other phase-field modeling studies of elemental materials and alloys \cite{Carter,LobWar02,NIST1}. In the light of the present results, it would be interesting to test for its existence in highly partitioning alloys with  interfaces that are structurally and chemically diffuse. High-energy boundaries in such materials appear to be the most likely candidates to observe coexistence of equilibrium states with different liquid film widths close to melting.

Secondly, we have found that a uniaxial stress perpendicular to the grain boundary plane is equivalent to a temperature change through a standard Clausius-Clapeyron relation. Therefore, for repulsive boundaries, a large tensile stress of magnitude comparable to a fraction of the yield stress should suffice to produce an observable increase of the liquid film width slightly below the melting point. This effect should be readily testable by molecular dynamics simulations.


\begin{acknowledgments} 
This work was supported by DOE grant DE-FG02-07ER46400. R.S. also acknowledges support of the German DFG grant SPP 1296 and the financial support from the industrial sponsors of ICAMS, ThyssenKrupp Steel AG, Salzgitter Mannesmann Forschung GmbH, Robert Bosch GmbH, Bayer Materials Science AG, Bayer Technology Services GmbH, Benteler AG and the state of North-Rhine-Westphalia for the later part of this work.
\end{acknowledgments} 

\appendix

\section{Equivalence of constrained and unconstrained formulations of the mechanical analog}
\label{eliminate}

Here we show explicitly that the elimination of the third field leads to the same result as keeping all independent fields and the additional constraint $\phi_1+\phi_2+\phi_3=1$.
We use a general free-energy density $f=f(\phi_1, \phi_2, \phi_3, \dot{\phi}_1, \dot{\phi}_2, \dot{\phi}_3)$ and the constraint $\phi_1+\phi_2+\phi_3=1$.
Then the stationary phase field equations are obtained from variation of the functional
\begin{equation}
\tilde{F} = \int dV \tilde{f},
\end{equation}
with $\tilde{f} = f- \lambda_0(x)(\phi_1+\phi_2+\phi_3-1)$.
Stationarity requires
\begin{equation}
\frac{\delta \tilde{F}}{\delta\phi_i}=0, \quad i=1,2,3. 
\end{equation}
From that we get the expression of the Lagrange multiplier
\begin{equation}
\lambda_0 = \frac{1}{3} \sum_{i=1}^3 \left( \frac{\partial f}{\partial\phi_i} - \frac{d}{dx}\frac{\partial f}{\partial\dot{\phi}_i} \right)
\end{equation}
and e.g. the first equation
\begin{equation} \label{eliminate::eq1}
2\frac{\partial f}{\partial \phi_1} - \frac{\partial f}{\partial \phi_2} - \frac{\partial f}{\partial \phi_3} = \frac{d}{dx} \left( 2\frac{\partial f}{\partial\dot{\phi}_1} - \frac{\partial f}{\partial\dot{\phi}_2} - \frac{\partial f}{\partial\dot{\phi}_3}\right).
\end{equation}
The generalized momenta are
\begin{equation}
p_i = \frac{\partial f}{\partial \dot{\phi}_i}.
\end{equation}
The Hamiltonian is
\begin{equation}
-H = F - p_1\dot{\phi}_i - p_2\dot{\phi}_2 - p_3\dot{\phi}_3.
\end{equation}

Now we can also eliminate the third field from the beginning, introducing a new free-energy density
\begin{equation}
\overline{f}(\phi_1, \phi_2, \dot{\phi}_1, \dot{\phi}_2) = f(\phi_1, \phi_2, 1- \phi_1-\phi_2, \dot{\phi}_1, \dot{\phi}_2, -\dot{\phi}_1-\dot{\phi}_2),
\end{equation}
and we have only two equations of motion
\begin{equation}
\frac{\partial \overline{f}}{\partial \phi_i} - \frac{d}{dx}\frac{\partial \overline{f}}{\partial \dot{\phi}_i} = 0,\quad i=1,2.
\end{equation}
They are explicitly
\begin{eqnarray}
\frac{\partial f}{\partial\phi_1} - \frac{\partial f}{\partial\phi_3} - \frac{d}{dx} \left( \frac{\partial f}{\partial\dot{\phi}_1} - \frac{\partial f}{\partial\dot{\phi}_3} \right) &=& 0, \\
\frac{\partial f}{\partial\phi_2} - \frac{\partial f}{\partial\phi_3} - \frac{d}{dx} \left( \frac{\partial f}{\partial\dot{\phi}_2} - \frac{\partial f}{\partial\dot{\phi}_3} \right) &=& 0.
\end{eqnarray}
Combining them gives us e.g. the same equation of motion (\ref{eliminate::eq1}) above.
We can also calculate the energy, and define momenta:
\begin{equation}
\overline{p}_i = \frac{\partial \overline{f}}{\partial\dot{\phi}_i} 
\end{equation}
Therefore we get
\begin{eqnarray}
\overline{p}_1 &=& \frac{\partial f}{\partial\dot{\phi}_1} - \frac{\partial f}{\partial\dot{\phi}_3},  \\
\overline{p}_2 &=& \frac{\partial f}{\partial\dot{\phi}_2} - \frac{\partial f}{\partial\dot{\phi}_3},
\end{eqnarray}
and the Hamiltonian
\begin{equation}
-\overline{H} = \overline{F} - \overline{p}_1\dot{\phi}_1 - \overline{p}_2\dot{\phi}_2 = -H,
\end{equation}
reduces to the same expression as before.
Here we used $\dot{\phi}_3=-\dot{\phi}_1-\dot{\phi}_2$.

\section{Numerical method}
\label{shooting}

Here we briefly explain the numerical shooting method used to solve the stationary phase-field equations, which have the form of a coupled ordinary differential equations (ODEs). For simplicity, we describe this method for the two-phase-field model of section \ref{attrrep} but the same method is applicable to the the multi-order parameter models after elimination of one field using the constraint $\phi_1+\phi_2+\phi_3=1$.

We integrate the ODEs starting in the left grain, i.e. for a large negative $x$ with the origin chosen midway between the grains. To find out the initial conditions for this integration, we linearize the phase field equations around the fixpoints, i.e.~$\psi=-1+\delta\psi$ and $\phi=-\delta\phi$.
Then the phase field equations (\ref{attrrep::eq4}) become to first order (for simplicity $h=\sigma=1$)
\begin{eqnarray*}
\frac{\partial^{2}\delta\phi}{\partial x^{2}} &=& (8+8\phi_{0}^{2})\delta\phi+6L\frac{T-T_M}{T_M}\frac{\delta\phi}{\phi_{0}^{2}} \\
\frac{\partial^{2}\delta\psi}{\partial x^{2}} &=& \frac{1}{\alpha}(8+8\phi_{0}^{2})\delta\psi
\end{eqnarray*}
and only have the exponential solutions
\begin{eqnarray*}
\psi &=& -1 + C_1 \exp [(8+8\phi_0^2)^{1/2}\alpha^{-1/2} x], \\
\phi &=& C_2 \exp \left[ \left( 8+8\phi_0^2+6\frac{L(T-T_M)}{T_M\phi_0^2} \right)^{1/2} x \right],
\end{eqnarray*}
which vanish at $x\to-\infty$, as required for a physically admissible solution;  
this requirement fixes one of the two integration constants for each second order equation. The remaining two constants $C_1$ and $C_2$ are used as adjustable parameters in the shooting method to fulfill the boundary conditions $d\phi/dx=0$ and $\psi=0$ at the midpoint between the two grains, which follow from the fact that $\phi$ and $\psi$ are symmetrical and antisymmetrical about this point, respectively. 
One of the shooting constants can be set to an arbitrary value since the problem is invariant under a translation along $x$, i.e. it just fixes the position of the origin.
Therefore, we integrate from a point far in the solid where the asymptotic analytical solutions are valid up to the point where $d\phi/dx=0$ is reached. We then use the remaining shooting parameter to fulfill the other boundary condition $\psi=0$.
The value of the liquid layer thickness can then be extracted by measuring twice the distance between the point where $\psi=-1/2$ and the endpoint of integration, since the profile is symmetric with respect  to the latter point.

For the multi-order parameter model, a similar strategy can be employed.
As definition of the liquid layer thickness we use here the distance between the points where the two solid fields cross the value $1/2$.

The stable branches can of course also be found by full relaxation according to Eq.~(\ref{multi::eq3}), but the above procedure is more accurate and efficient also in these cases.

\section{Double-obstacle potential}
\label{obstacle}

Instead of the multi-well a multi-obstacle potential is often used for phase field simulations \cite{Steinbach09}, and we briefly investigate its behavior concerning short-range interface interactions here.
We refrain from a full analytical and numerical treatment and discuss for simplicity only a model with a single order parameter.

\begin{figure}
\begin{center}
\includegraphics[width=8.5cm]{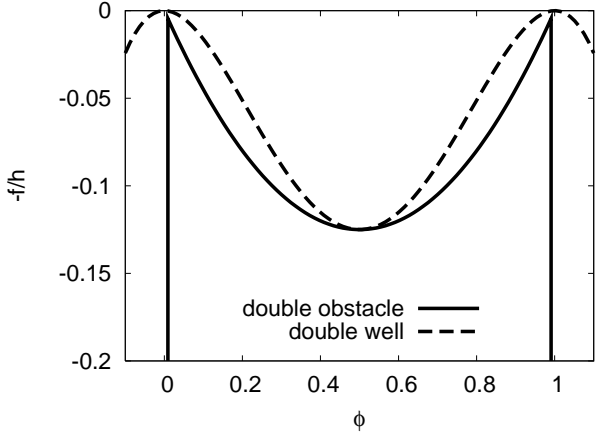}
\caption{Mechanical potential energy versus phase-field $\phi$ (negative of the free-energy density) for the double-obstacle and double-well potentials.}
\label{obstacle::fig1}
\end{center}
\end{figure}

The main difference between the double-well and the double-obstacle potential is that the latter is defined to be infinite outside the the physical regime $0\neq \phi \neq 1$, and this is sketched in Fig.~\ref{obstacle::fig1}.
Furthermore, the potential has a finite slope at the end points $\phi=0$ and $\phi=1$.
A typical choice is 
\begin{equation}
f_{DO} = \left\{ \
\begin{array}{cc}
h\phi(1-\phi) & \mathrm{for }\, 0\neq \phi \neq 1 \\
\infty & \mathrm{else}
\end{array}
\right.
\end{equation}
instead of the double well potential
\begin{equation}
f_{dw} = 2h\phi^2(1-\phi)^2.
\end{equation}
\begin{figure}
\begin{center}
\includegraphics[width=8cm]{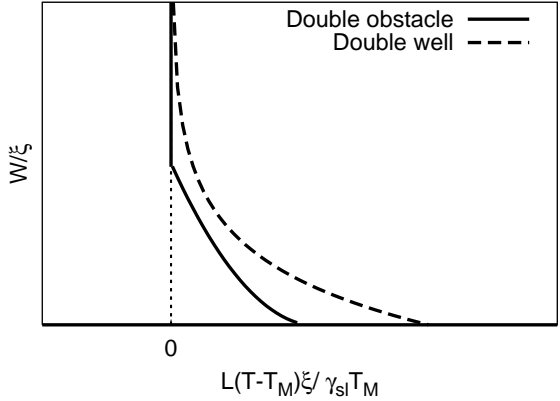}
\caption{Sketch of the liquid layer thickness as function of temperature for the double-well and double-obstacle potential in the framework of a single-order parameter model. The precise location of the curves depends on the model and the definition of the melt layer thickness.}
\label{obstacle::fig2}
\end{center}
\end{figure}
Again, we use a standard kinetic term of the type
\begin{equation}
f_k = \sigma (\nabla\phi)^2.
\end{equation}
The central point is now that the double-obstable potential provides stationary interface solutions which have only a finite support, i.e.~the phase field differs from the trivial values $\phi=0$ or $\phi=1$ only in a {\em finite} region.
For the double-well potential, the phase field approaches these limiting values only exponentially.
For the above choice of free-energy contributions, the stationary interface solution $\phi(x)$ is given by  
\begin{equation}
 \left\{
\begin{array}{cc}
0 & \mbox{if } x-x_0 < -\xi \pi/2 \\
\frac{1}{2}\left( 1+\sin [(x-x_0)/\xi]\right) & \mbox{if } -\xi \pi/2 < x-x_0 < \xi \pi/2 \\
1 & \mbox{else}
\end{array}
\right.
\end{equation}
where $x_0$ is the interface position for this one-dimensional solution and $\xi=(\sigma/h)^{1/2}$ is a measure for the interface thickness, as before.

If we repeat the mechanical analog of Fig. \ref{onepf} for the double-obstacle potential, it is clear that the
particle trajectory can only exist for $T>T_M$ since a turning point is still present. Therefore the interaction is still attractive. The main difference is that the potential does not have zero slope near the liquid maximum. The potential has a finite slope that does not change as 
$T-T_M\rightarrow 0^+$. This slope implies that the particle has a constant negative acceleration at the turning point even in this limit. Therefore, it spends a finite amount of time near the turning point and $W$ does not diverge in this limit.
Instead, it reaches a maximum value as shown schematically in Fig. \ref{obstacle::fig2}. Exactly at $T=T_M$, liquid films can exist for any $W$ larger than this maximum since the interaction between interfaces becomes strictly zero. This also implies that the disjoining potential vanishes at a finite $W$ for the double-obstacle potential.


\end{document}